\documentclass{aa}  

\usepackage{graphicx}
\usepackage{txfonts}
\usepackage[english]{babel}
\usepackage{amssymb}
\usepackage{amsmath}
\usepackage{mathrsfs}
\usepackage{latexsym}
\usepackage{longtable}
\usepackage{multirow}
\usepackage{epsf}
\usepackage{color}
\usepackage{float}
\usepackage{enumerate}
\usepackage{txfonts}
\usepackage{natbib}
\usepackage{comment}
\usepackage{capt-of}
\usepackage{rotating}
\usepackage{ulem}

\newcommand{ \s}{$\sim$}
\newcommand{\dg}{$^\circ$}
\newcommand{\as}{\arcsec{}}

\newcommand{\hd}{HD\,143006 }

\begin{document} 

   \title{Shadows and asymmetries in the T Tauri disk HD\,143006: \\ Evidence for a misaligned inner disk
\thanks{Based on observations performed with SPHERE/VLT under program ID 097.C-0902(A) and 095.C-0693(A).}}

\author{M.\,Benisty\inst{1,2}
\and A.\,Juh\'asz\inst{3}
\and S.\,Facchini\inst{4}
\and P.\,Pinilla\inst{5}
\and J.\,de~Boer\inst{6}
\and L.\,M.\,P\'erez\inst{7}
\and M.\,Keppler\inst{8}
\and G.\,Muro-Arena\inst{9}
\and M.\,Villenave\inst{10,2}
\and S.\,Andrews\inst{11}
\and C.\,Dominik\inst{9}
\and C.\,P.\,Dullemond\inst{12}
\and A.\,Gallenne\inst{10}
\and A.\,Garufi\inst{13}
\and C.\,Ginski\inst{6,9}
\and A.\,Isella\inst{14}}

\institute{Unidad Mixta Internacional Franco-Chilena de Astronom\'{i}a (CNRS, UMI 3386), Departamento de Astronom\'{i}a, Universidad de
Chile, Camino El Observatorio 1515, Las Condes, Santiago, Chile
\email{Myriam.Benisty@univ-grenoble-alpes.fr}
\and Univ. Grenoble Alpes, CNRS, IPAG, 38000 Grenoble, France. 
\and Institute of Astronomy, Madingley Road, Cambridge CB3 OHA, UK
\and Max-Planck-Institut f\"ur Extraterrestrische Physik, Giessenbachstrasse 1, 85748 Garching, Germany
\and Department of Astronomy/Steward Observatory, The University of Arizona, 933 North Cherry Avenue, Tucson, AZ 85721, USA
\and Leiden Observatory, Leiden University, P.O. Box 9513, 2300 RA Leiden, The Netherlands
\and Universidad de Chile, Departamento de Astronoma, Camino El Observatorio 1515, Las Condes, Santiago, Chile
\and Max Planck Institute for Astronomy, K\"{o}nigstuhl 17, 69117 Heidelberg, Germany
\and Anton Pannekoek Institute for Astronomy, University of Amsterdam, Science Park 904,1098XH Amsterdam, The Netherlands
\and European Southern Observatory, Alonso de C\'ordova 3107, Vitacura, Casilla 19001, Santiago, Chile
\and Harvard-Smithsonian Center for Astrophysics, 60 Garden Street, Cambridge, MA 02138
\and Zentrum f\"ur Astronomie, Heidelberg University, Albert-Ueberle-Strasse 2, D-69120 Heidelberg, Germany
\and INAF, Osservatorio Astrofisico di Arcetri, Largo Enrico Fermi 5, I-50125 Firenze, Italy
\and Department of Physics and Astronomy, Rice University, 6100 Main Street, Houston, TX, 77005, USA
}

\date{}

  \abstract
   {
   While planet formation is thought to occur early in the history of a protoplanetary disk, the presence of planets embedded in disks, or of other processes driving disk evolution, might be traced from their imprints on the disk structure. }
   {We study the morphology of the disk around the T Tauri star HD\,143006, located in the \s5-11\,Myr-old Upper Sco region, and we look for signatures of the mechanisms driving its evolution.}
   {We observed \hd in polarized scattered light with VLT/SPHERE at near-infrared ($J$-band, 1.2\,$\mu$m) wavelengths, reaching an angular resolution of \s0.037\as (\s6\,au). We obtained two datasets, one with a 145\,mas diameter coronagraph, and the other without, enabling us to probe the disk structure down to an angular separation of \s0.06\as (\s10\,au).}
   {In our observations, the disk of \hd is clearly resolved up to \s0.5\arcsec{} and shows a clear large-scale asymmetry with the eastern side brighter than the western side. We detect a number of additional features, including two gaps and a ring. The ring shows an overbrightness at a position angle (PA) of \s140\dg, extending over a range in position angle of \s60\dg, and two narrow dark regions. 
The two narrow dark lanes and the overall large-scale asymmetry are indicative of shadowing effects, likely due to a misaligned inner disk. 
We demonstrate the remarkable resemblance between the scattered light image of HD\,143006 and a model prediction of a warped disk due to an inclined binary companion. The warped disk model, based on the hydrodynamic simulations combined with three-dimensional radiative transfer calculations, reproduces all major morphological features. However, it does not account for the observed overbrightness at PA\s140\dg. }
{Shadows have been detected in several protoplanetary disks, suggesting that misalignment in disks is not uncommon. However, the origin of the misalignment is not clear.  As-yet-undetected stellar or massive planetary companions could be responsible for them, and naturally account for the presence of depleted inner cavities.} 
\keywords{Protoplanetary disks -- Techniques: polarimetric -- Radiative transfer -- Scattering}

\titlerunning{A warp in the inner disk around the T Tauri star HD\,143006}
\authorrunning{Benisty et al.}
\maketitle

\section{Introduction}
\label{sec:introduction}

High angular resolution observations of protoplanetary disks show a wide diversity of structures on different scales. Sub-millimeter (mm) observations show radial structure, for example, bright and dark rings \citep[e.g.,][]{2015ApJ...808L...3A,andrews2016,2017A&A...600A..72F,dipierro2018}, and azimuthal asymmetries (sometimes called 'horseshoe') where the dust continuum emission is much stronger than in the surrounding background disk \citep[e.g.,][]{2013Sci...340.1199V,2013Natur.493..191C}. Atacama Large Millimeter Array (ALMA) observations at high angular resolution revealed that most, if not all, of the protoplanetary disks are not smooth, and often show  ring-like features \citep[][]{isella2016, fedele2018, cieza2017}, suggesting that these features trace universal processes.  One of the most interesting scenarios is that they are due to interactions between the planet-forming disk and embedded proto-planets \citep[e.g.,][]{dipierro2015,rosotti2016}. However, other physical mechanisms, such as ice lines, or dead zones, have also been invoked to explain such observations \citep[e.g.,][]{flock2015,zhang2015,bethune2016,okuzumi2016,pinilla2016,pinilla2017}.

Scattered light images show an even more diverse spectrum of structures, including cavities \citep[e.g.,][]{pinilla2015}, rings and gaps, sometimes colocated with sub-mm counterparts \citep[e.g.,][]{vanboekel2017,2017arXiv171006485P,avenhaus2018}, and multiple spirals \citep[e.g.,][]{garufi2013,benisty2015}.  While the continuum mm emission traces the cold disk midplane, scattered light observations trace the small dust particles in the disk upper layers that are directly irradiated by the young star, and hence also depend on the illumination pattern. Therefore, geometrical variations, such as scale-height perturbations due to temperature variations  \citep[e.g.,][]{juhasz2015}, or due to a warp, can create strong azimuthal asymmetries in scattered light \citep[e.g.,][]{juhasz2017,facchini2018}. 

Scattered light images of transition disks (disks with an inner dust depleted cavity) show clear evidence for misaligned inner regions, either with narrow shadow lanes in the outer disk  \citep{marino2015,pinilla2015,stolker2016,benisty2017,casassus2018}, or with low amplitude azimuthal variations observed over time  \citep{debes2017}.  Additional evidence for warps in disks are also found in sub-mm observations, through the kinematics of gas lines \citep{rosenfeld2012,casassus2015,brinch2016,walsh2017,loomis2017,boehler2018}. Dipper stars provide another example of potentially warped inner disks \citep[e.g.,][]{cody2014, bodman2017}. In that case, the warp is thought to be due to a strong dipolar stellar magnetic field that is misaligned with respect to the disk midplane and forces the innermost disk to tilt out of the plane \cite[e.g., AA\,Tau,][]{bouvier2007}. The dipper light curves show clear dimming events usually interpreted as the signature of material from a very inclined disk that repeatedly blocks the line of sight. However, recent imaging  of moderately inclined outer disks in some dipper stars suggested a strong misalignment between the inner and outer disk regions \citep{ansdell2016,loomis2017}.  Such a misalignment can be very large, up to \s70\dg\,\citep[e.g.,][]{marino2015,benisty2017, min2017}, and in some cases the resulting shadows can lead to a significant cooling of the outer disk material \citep{casassus2018}. 

In addition to the effect of a strongly inclined stellar magnetic field, warps can result from the gravitational interaction of a massive companion with the disk. If the orbit of the companion is significantly misaligned with respect to the disk midplane, the disk can break, which leads to a significant misalignment between the inner and outer regions of the disk \citep{nixon2012b,facchini2013,dougan2015,facchini2018}. Other mechanisms, such as secular interactions and  precessional resonances \citep{lubow2016,owen2017}, can also strongly tilt the inner disk.  Depending on the location of the companion, these scenarios can lead to a misaligned circumprimary or circumbinary disk.  In the first case, such a companion would naturally create a dust depleted inner cavity, as found in transition disks, and explain most of the properties of some of these disks  \citep[e.g., HD\,142527,][]{price2018}. So far, except in HD\,142527 and PDS\,70, the putative companion inside the cavity of transition disks remains to be clearly detected \citep{biller2012,keppler2018}. 



The focus of this study is the protoplanetary disk around \hd(also, 2MASS J15583692-2257153). \hd is a G7 T Tauri star with the following stellar parameters (updated after the Gaia Data Release 2): T$_{\rm{eff}}=5880$\,K, L$_{*}$=4.58\,L$_{\odot}$, and M$_{*}$=1.5$M_{\odot}$ \citep[][Garufi et al. in prep]{salyk2013}. 
It is located at a distance of 166$\pm$4\,pc \citep{gaia2018} and belongs to the Upper Sco star-forming region, which is rather old  \citep[5-11 Myr; ][]{preibisch2002,pecaut2012} compared to the typical timescales for disk evolution. \citet{barenfeld2016} observed $\sim$100 disk-host candidates in Upper Sco with ALMA, including HD\,143006, and showed that, not only does  Upper\,Sco have a lower fraction of disk-host stars than younger regions (such as Taurus or Lupus), but also that these disks have a lower dust mass to stellar mass ratio. The disk around \hd(hereafter only referred to as HD\,143006) is resolved at 0.88\,mm with a $\sim 0.35\arcsec{} \times0.3$\arcsec{} beam, and its continuum map shows a centrally depleted large cavity \citep[$\sim$84\,au,][]{pinilla2018b}, and a low-contrast brightness asymmetry with an enhanced emission (by a factor $\lesssim$2) in the south-east \citep{barenfeld2016}. The innermost region, however, remains dust- and gas-rich as indicated by the high near infrared (IR) excess (\s21\%, Garufi et al. submitted), and by near IR  interferometric observations that resolve hot dust at the sub-au scale \citep{lazareff2017}. A relatively strong H$_{\alpha}$ emission line translates to a mass accretion rate of $\sim2\times10^{-8}\,M_{\odot}$\,yr$^{-1}$ \citep{rigliaco2015}.  

In this paper we present the first scattered light observations of \hd obtained with the Very Large Telescope (VLT) Spectro-Polarimetric High-contrast Exoplanet REsearch (SPHERE) instrument. Our observations trace the small (sub- and micron-sized) dust grains, well coupled to the gas, in a tenuous surface layer of the disk and show a number of features that share striking similarities with the predictions of a warped disk model. 
Our paper is organized as follows: in Sect.\,\ref{sec:datared} we present our observations and the data reduction; in Sect.\,\ref{sec:images} we describe the scattered light images, in Sect.\,\ref{sec:modeling}, the hydrodynamical simulations and radiative transfer predictions, and in Sect.\,\ref{sec:discussion} we discuss our findings. 

\begin{figure*}
\center
\begin{tabular}{ccc}
\includegraphics[width=0.33\textwidth]{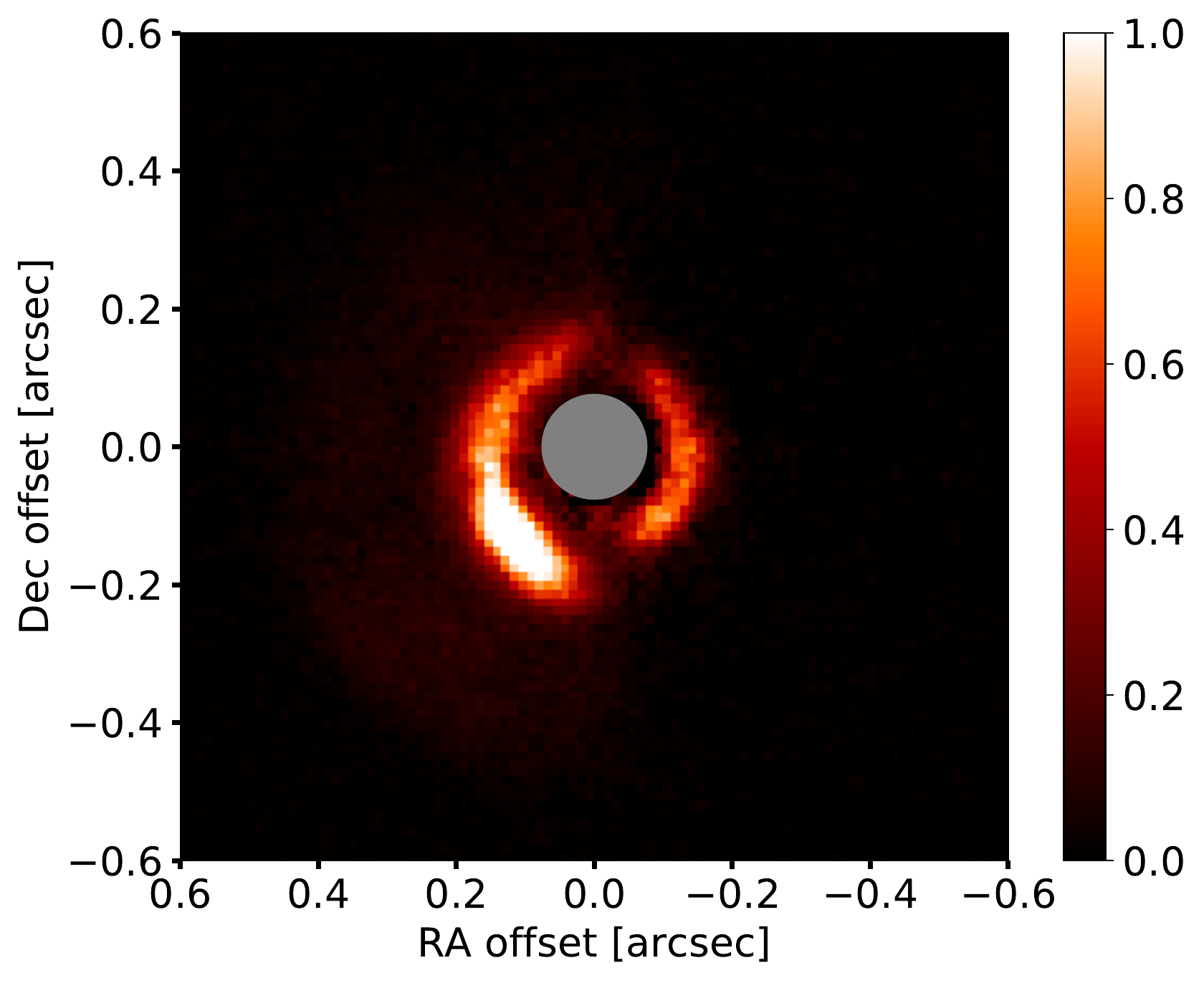} & 
\includegraphics[width=0.33\textwidth]{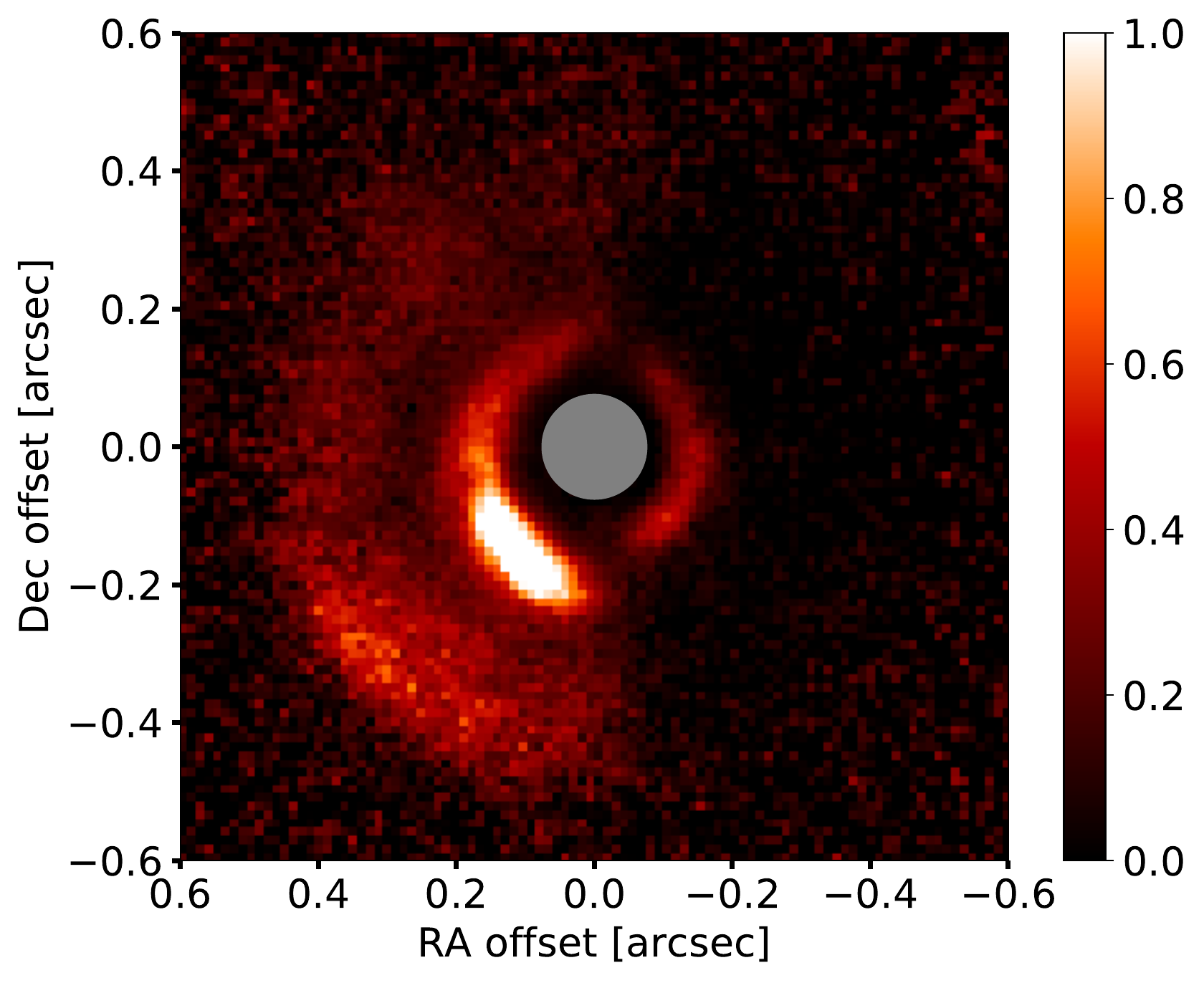} & 
\includegraphics[width=0.33\textwidth]{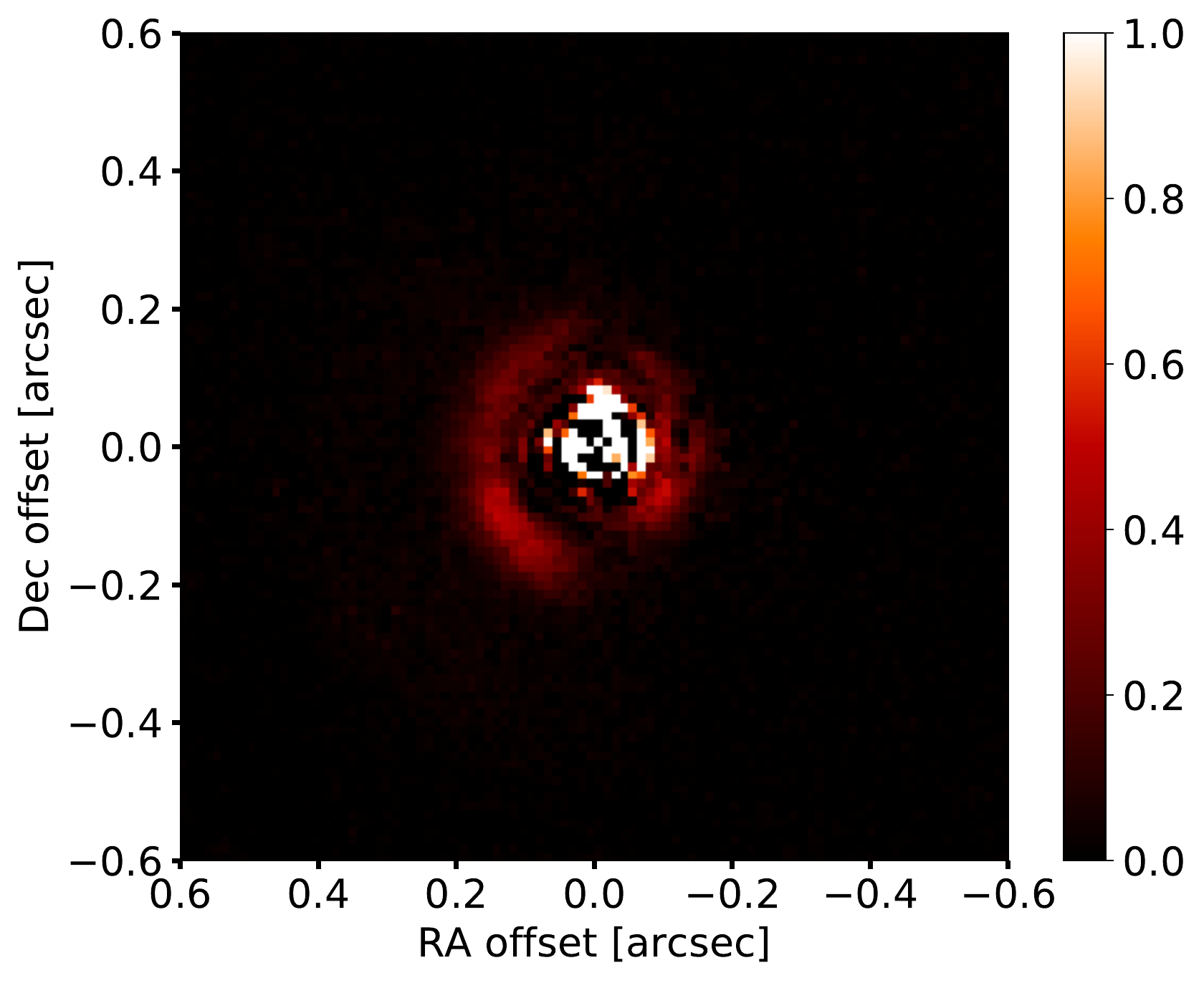}
\end{tabular}
\caption{
Left: $J$-band polarized intensity image $Q_{\phi}$ obtained with a coronagraph. The region masked by the coronagraph is indicated by the gray circle. Middle: Same image but scaled by  $r^2$ to account for the drop-off in stellar illumination and enhance the visibility of faint outer disk features. Right: additional $Q_{\phi}$ image obtained without coronagraph. The innermost pixels (\s60\,mas) are saturated. The color scales are arbitrary. 
North is up, east points toward the left.} 
\label{fig:obsQphi}
\end{figure*}

\begin{figure*}
\center
\begin{tabular}{cc}
\includegraphics[width=0.45\textwidth]{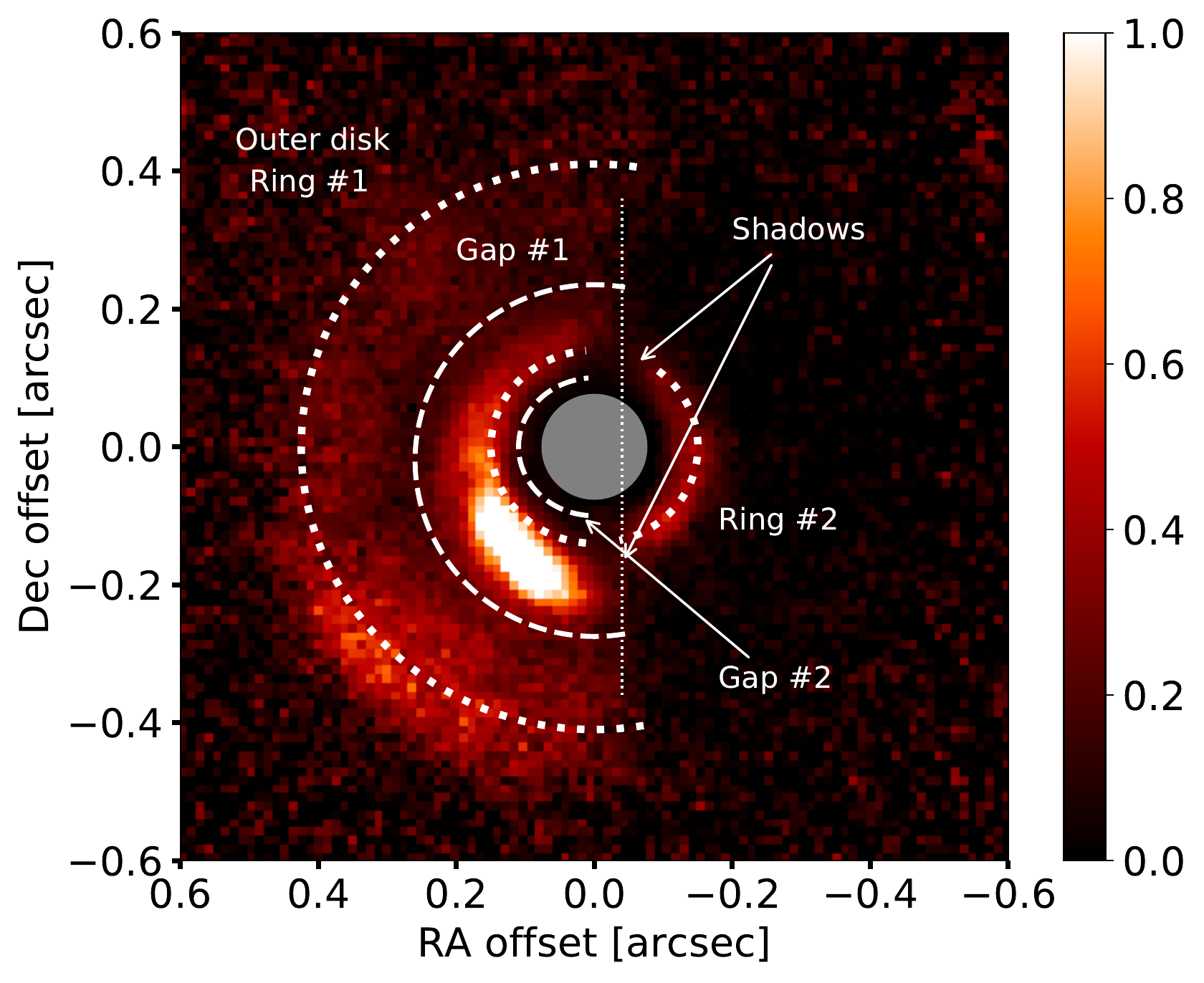} & 
\includegraphics[width=0.465\textwidth]{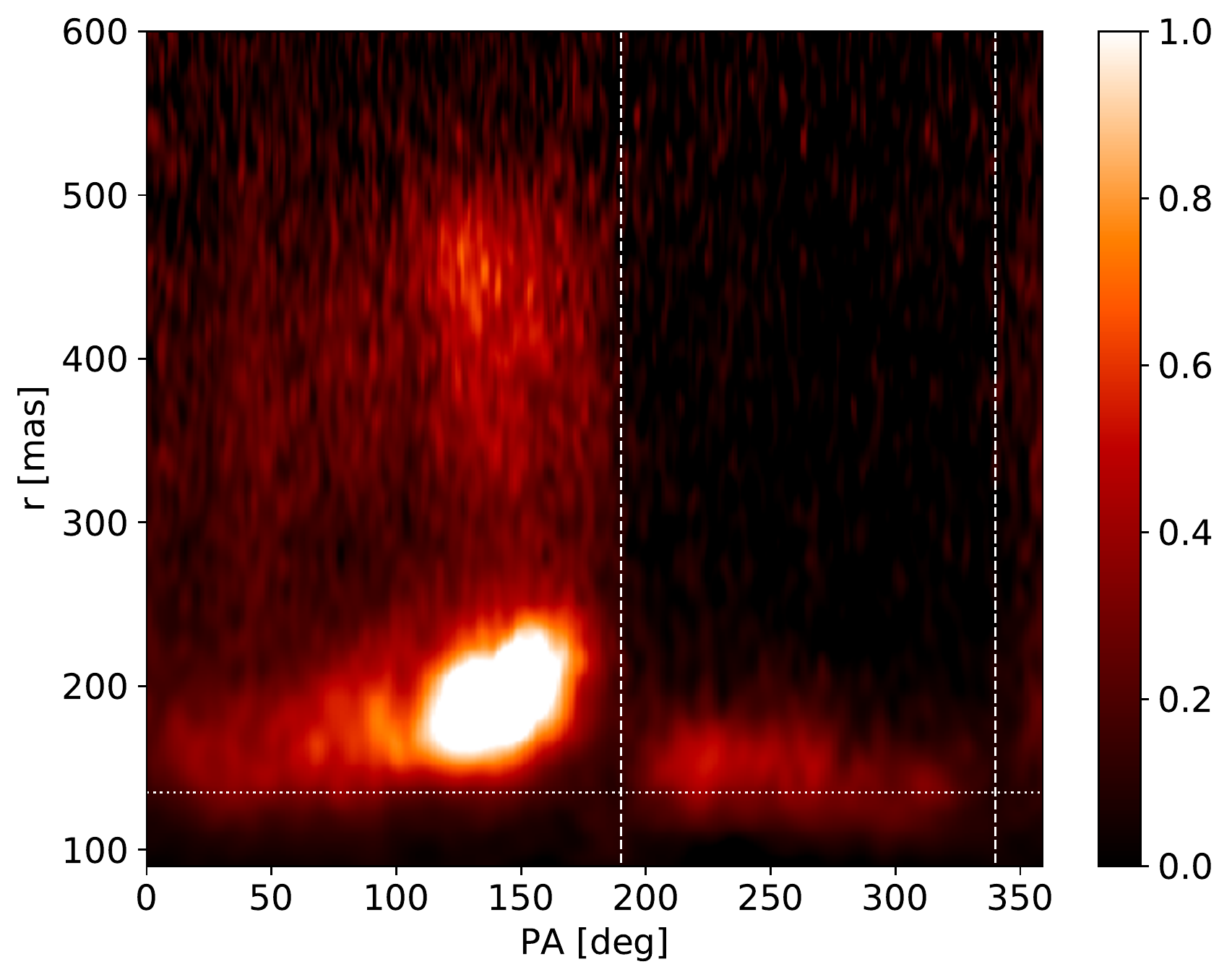}
\end{tabular}
\caption{Left: $J$-band coronagraphic polarized intensity image $Q_{\phi}$, scaled by  $r^2$, with annotations. Right: Polar mapping of the  $r^2$-scaled image, deprojected with i=17\dg\,and PA=170\dg. The two vertical lines indicate the locations of the two shadow lanes (at PA\s190\dg\,and 340\dg), while the horizontal dashed line indicates a radius of 0.135\arcsec{}.}
\label{fig:obsQphiannotatedpolarmap}
\end{figure*}

\section{Observations and data reduction}
\label{sec:datared}
\subsection{SPHERE imaging}
We obtained observations at the Very Large Telescope located at Cerro Paranal, Chile, on 2016 June 30, using the SPHERE instrument \citep{beuzit2008}. SPHERE features an extreme adaptive-optics (AO) system \citep{fusco2006,petit2014,sauvage2014} that feeds three science channels and enables high angular resolution and high-contrast imaging at optical (visible and near-infrared) wavelengths. \hd was observed with the polarimetric imaging mode of the InfraRed Dual-band Imager and Spectrograph \citep[IRDIS;][]{dohlen2008,langlois2014}, in the $J$ band  ($\lambda_0$=1.258, $\Delta\lambda$=0.197\,$\mu$m). IRDIS has a plate scale in the $J$ band of 12.26\,mas per pixel \citep[][]{maire2016}. In addition, to enhance the detection of outer disk features, we used a 145\,mas-diameter coronagraphic focal mask (N\_ALC\_YJ\_S, with an inner working angle of 0.08\arcsec{}, \citealt{martinez2009,carbillet2011}) for one dataset, but removed it for the second dataset, to enable the observation of inner regions otherwise covered by the coronagraph. \hd was observed for \s35 minutes on-source with the coronagraph, and approximately five minutes on sources without coronagraph, with a seeing of 0.6\as-0.8\as and 4-5 milliseconds of coherence time. The analysis of the point spread function (PSF) that is estimated from a non-coronagraphic FLUX (short, non-saturated images of the star outside the masked region) measurement shows that the observations reach a 37\,mas$\times$37\,mas resolution and a Strehl ratio of 51\%. The five inner pixels of the non-coronagraphic polarimetric image (within a radius of \s60 mas) are saturated. 

With polarimetric differential imaging \citep[PDI; e.g.,][]{kuhn2001,apai2004} one measures the linear polarization of the light scattered by dust grains in the disk. This technique enables us to efficiently remove the unpolarized stellar contribution and to image with high contrast the outer disk from which we detect polarized scattered light. The instrument splits the beam into two orthogonal polarization states, and a half-wave plate (HWP) is set to four positions shifted by 22.5$^\circ$ in order to construct a set of linear Stokes images. We reduce the data according to the double difference method \citep{kuhn2001}, and derive the Stokes parameters $Q$ and $U$. Assuming only one scattering event for each photon, the scattered light from a protoplanetary disk, at low inclination angle\footnote{In inclined disks, multiple scattering effects lead to a strong signal in $U_{\phi}$ \citep[e.g., T\,Cha:][]{pohl2017}.}, is expected to be linearly polarized in the azimuthal direction. We therefore describe the Stokes parameters in polar coordinates \citep[$Q_{\phi}$, $U_{\phi}$;][]{schmid2006,avenhaus2014}, as
$$Q_{\phi} = +Q \cos(2\phi) + U \sin(2\phi)$$
$$U_{\phi} = -Q \sin(2\phi) + U \cos(2\phi)$$
\noindent with $\phi$ the position angle of each pixel (x, y) with respect to the star location. The polarized flux along the azimuthal direction appears as a positive signal in the $Q_{\phi}$ image, while radial polarization will lead to negative $Q_{\phi}$.  If there is only azimuthal and/or radial polarization, the $U_{\phi}$ image contains no disk signal and can be used as an estimate of the residual noise in the $Q_{\phi}$ image \citep{schmid2006}. We correct for the instrumental polarization by minimizing $U_{\phi}$ and subtracting scaled versions of the total intensity frame from the Stokes Q and U frames \citep{canovas2011}. The final images were corrected for the true north \citep[by rotating them by 1.775\dg\, in the counterclockwise direction,][]{maire2016}. The resulting images are shown in Fig.\,\ref{fig:obsQphi}. 

\subsection{ALMA observations}
\hd was observed by ALMA in Cycle\,2 (2013.1.00395.S) with a synthesized beam of 0.35\arcsec$\times$0.30\arcsec, PA=-73\dg. For more details on the data reduction and calibration, we refer the reader to \citet{barenfeld2016}, who first presented the data. In this paper, we will only discuss the optically thick $^{12}$CO\,J=3-2 observations that trace the surface layers as our scattered light observations.

\begin{figure*}
\center
\begin{tabular}{cc}
\includegraphics[width=0.46\textwidth]{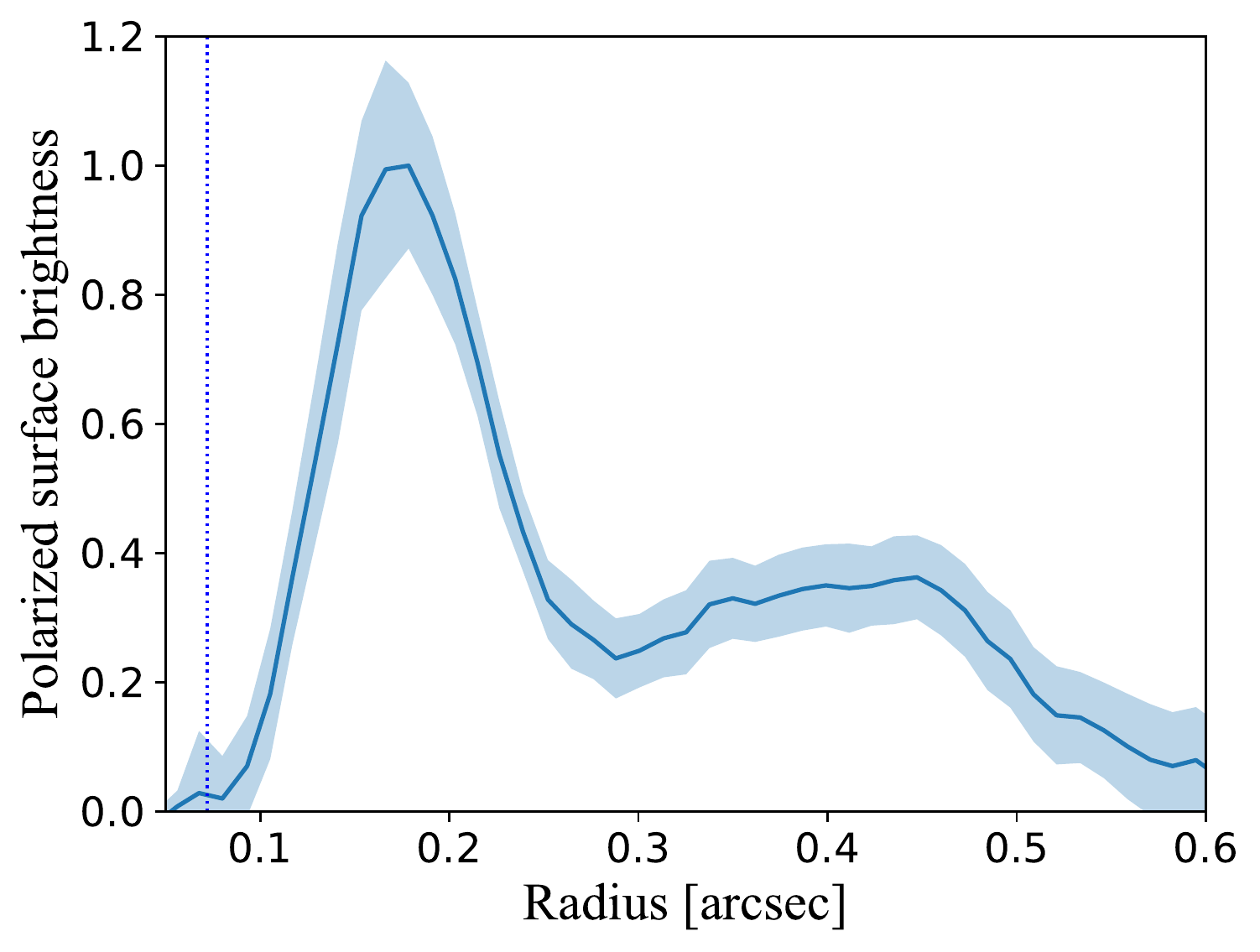} & 
\includegraphics[width=0.46\textwidth]{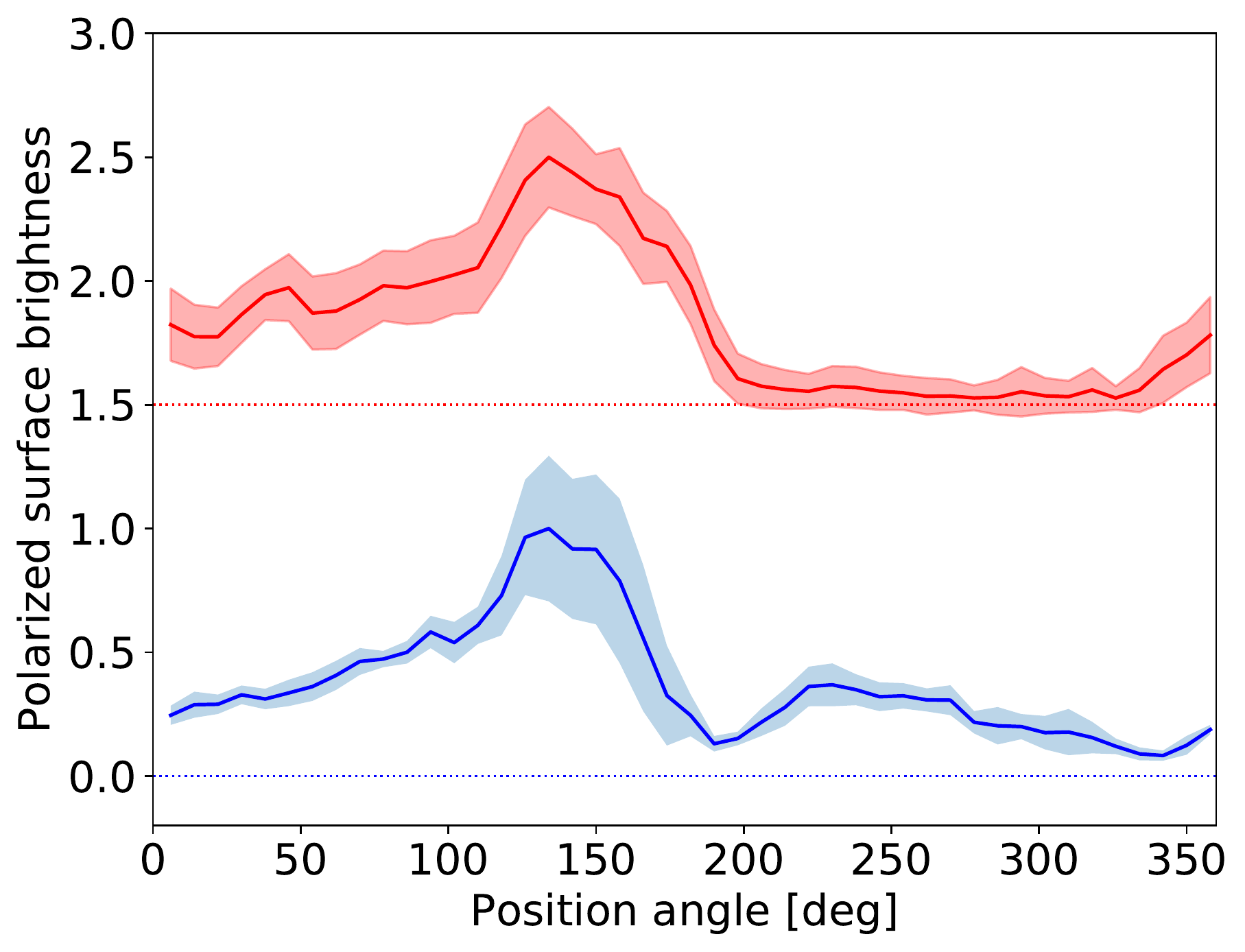}
 \end{tabular}
\caption{Left: Radial profile of the r$^{2}$-scaled Q$_{\phi}$ image after deprojection (with i=17\dg\,and PA=170\dg) and azimuthal averaging. The plot  indicates the presence of an inner cavity, that extends beyond the coronagraph (defined with the vertical dotted line).  The error bars are defined as the standard deviation of Q$_{\phi}$ in the corresponding bin, divided by the square root of the number of PSF contained in the bin. Right: Azimuthal profile of the same image after deprojection and averaging across the Ring\,\#2 and Ring\,\#1 widths (0.12\as-0.17\as, blue curve; 0.4\as-0.48\as, red curve, respectively). Each curve is normalized to its maximum value, and the red profile is offset by $1.5$ for clarity. The error bars are indicated with the shaded region, defined as the standard deviation of Q$_{\phi}$ in the corresponding bin, which reflects the large variations of the surface brightness. }
\label{fig:radazicuts}
\end{figure*}

\section{Scattered light images} 
\label{sec:images}
Our scattered light images, with and without coronagraph, are shown in Fig.\,\ref{fig:obsQphi}, and show distinct features that are annotated in Fig.\,\ref{fig:obsQphiannotatedpolarmap}, left panel.  These features are more evident when the image is scaled by $r^2$ to account for the drop of the stellar illumination with radius. Figure\,\ref{fig:radazicuts} provides radial and azimuthal profiles.

As indicated in the left panel of Fig.\,\ref{fig:obsQphiannotatedpolarmap}, we detect the following features, from outside in: 

\textbf{(a)} an outer disk extending from \s0.3\as\,to \s0.5\as\,(\s50 to \s83\,au) in scattered light. We refer to this region as Ring\,\#1. It shows a clear azimuthal asymmetry, and is only detected between 0 and 185\dg. 

We measure a ratio of (radially averaged) polarized surface brightness of \s0.4/1 between position angles (PA) \s50\dg\,and 140\dg\,(where the overbrightness lies), respectively. The ratio of the brightness along east/west along PA\s50\dg\,(and 230\dg) is \s1/0.05 (see Fig.\,\ref{fig:radazicuts}). 

\textbf{(b)} a region with less polarized signal than the surrounding disk, between \s0.24\as\,to \s0.3\as\,(\s40 to 50\,au), which we call Gap\,\#1. We note that this region is not devoid of scattered light signal at our angular resolution (see Fig.\,\ref{fig:radazicuts}, left panel). We measure a ratio of polarized surface brightness, after azimuthal averaging, of \s70\% between radii of 0.3\as and 0.4\as. 

\textbf{(c)} a ring-like feature (Ring\,\#2) between \s0.11\as\,to \s0.18\as\,(\s18 to 30\,au), that presents a strong azimuthal asymmetry. At PAs of \s190\dg~and 340\dg, two dark regions that we will refer to as shadows, can be seen, and at PAs between \s110\dg-170\dg, we observe an overbrightness. This range of PAs is the same as the one over which the outer disk is the brightest.  We also note that the peak signal of that ring is at different separations on the west and east side. 

Along Ring\,\#2, we measure a ratio of brightness of \s0.3/1 between PA\s50\dg\,and 140\dg, respectively, and a similar brightness along east/west. Finally, we measure a radially averaged polarized surface brightness of \s7 and \s12\% of the max brightness at the shadows' locations  (see Fig.\,\ref{fig:radazicuts}). 

\textbf{(d)} a dark region, called Gap\,\#2, inside 0.11\as\,(\s18\,au). The non-coronagraphic image (Fig.\,\ref{fig:obsQphi}, right) provides a better view of that region and supports the detection of a gap, which we marginally detect right outside the limits set by the saturated pixels (\s0.06\arcsec{}, i.e., \s10\,au). The presence of  these saturated inner pixels prevents us from determining the inner edge, if any, of this gap, and whether there are other rings inside 10\,au. It also prevents a direct detection of an inner disk inside this area. However, the near-IR excess observed in the spectral energy distribution (SED) of \hd indicates the presence of hot dust at the sublimation radius, which was also spatially resolved with near-IR interferometry  \citep[at \s0.1\,au,][]{lazareff2017}.

All the features are  very apparent in the polar mapping of the  $r^2$-scaled image. Figure\,\ref{fig:obsQphiannotatedpolarmap}, right,  presents such an image, obtained after  deprojection using i=17\dg\,and PA=170\dg\,(see Sect.\,\ref{sec:almaimaging}). It clearly shows an east/west asymmetry, as well as the bright area between PA\s of 110\dg\,to 170\dg. This overbrightness on Ring\,\#2 does not appear to be co-radial and has some contribution from larger radii. The overbrightness along this range of PA is also evident in the outer disk. The two shadow lanes, the inner gap inside \s0.11\as\,(Gap \#2), and the outer disk (Ring\,\#1 and Gap\,\#1) are also clearly visible. 

The presence of an inner gap (Gap \#2), depleted in small dust grains, as well as of a second, much shallower gap (Gap\,\#1), clearly appears in the radial profile of the $r^2$-scaled Q$_{\phi}$ image (Fig.\,\ref{fig:radazicuts}, left). This plot is produced after deprojection and azimuthal averaging. The right panel presents the azimuthal profiles of Rings \#1 and \#2 after deprojection and an average over a width of [0.13\as-0.19\as] and [0.32\as-0.47\as], respectively.  The large azimuthal brightness asymmetries are clearly visible.  

The images shown in Figs.\,\ref{fig:obsQphi} and \ref{fig:obsQphiannotatedpolarmap}, left, are not deprojected. The middle panel of Fig.\,\ref{fig:obsQphi}, and both panels of Fig.\,\ref{fig:obsQphiannotatedpolarmap} are shown with $r^2$ scaling to account for the drop-off in stellar illumination and enhance the visibility of faint outer features. This procedure does not take into account the effects of inclination and PA and of the non-planarity of the surface layers that scatter the stellar light  \citep[see such a method in][]{stolker2016b}. Considering the low inclination of the object (as determined in Sect.\,\ref{sec:almaimaging}), these effects are expected to be small and should not dramatically affect the shape of the features described in this section. 

\section{Modeling}
\label{sec:modeling}

\subsection{Inner and outer disk misalignments}
\label{sec:almaimaging}
The characteristics of shadows observed in scattered light images of disks have been well reproduced with a significant misalignment (up to \s70\dg) between inner and outer disk regions \citep[e.g.,][]{min2017}. To investigate if this could apply to HD\,143006, we discuss in this section the inclination and PA values inferred for HD\,143006 using various tracers of the inner and outer disk.

\begin{figure*}
\includegraphics[width=1.0\textwidth]{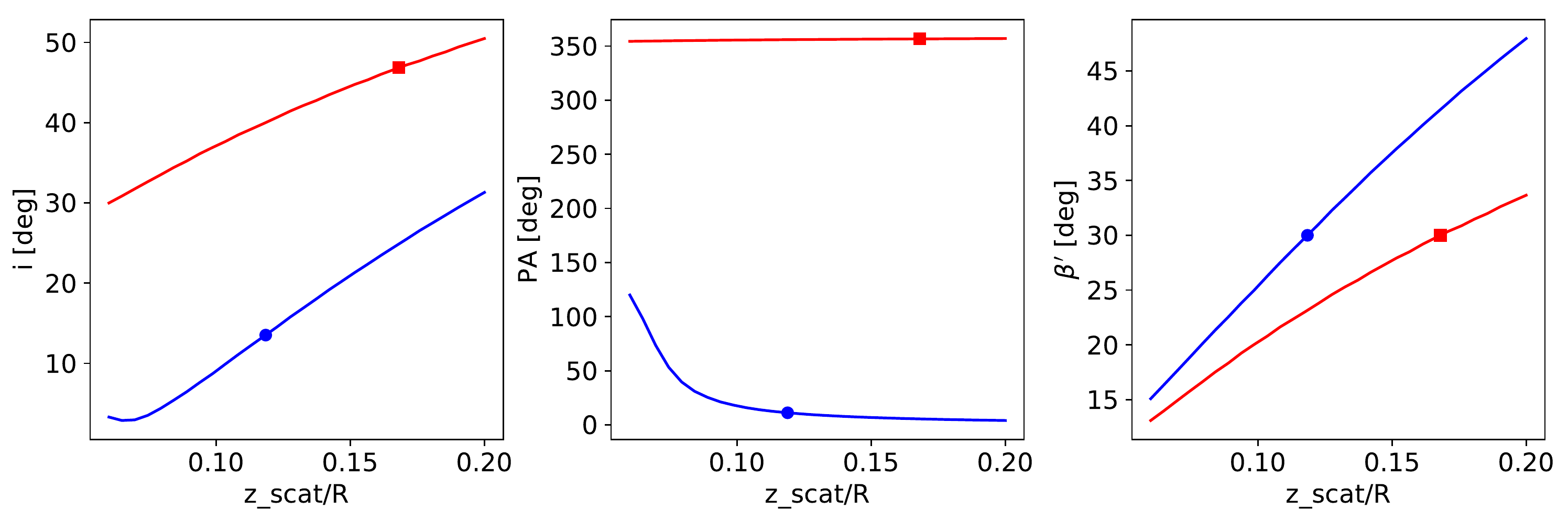}
\caption{Solutions for the orientation of the inner disk, based on the location of the shadows and assuming i=17\dg\,for the outer disk, shown against z$_{\rm{scat}}$/R. Left: Inner disk inclination; Middle: PA of the inner disk; Right: Misalignment angle, $\beta^{\prime}$, between the inner and outer disk. The colors (red and blue) correspond to the two families of solutions depending on the near side of the outer disk (blue, for an outer disk PA of 170\dg\,with the near side in the east; red, for an outer disk PA of 350\dg\,with the near side in the west). The  circles and squares  indicate a misalignment of 30\dg.}
\label{fig:misalignments}
\end{figure*}

\paragraph{Outer disk.}
Among other transition disks, \citet{pinilla2018a} modeled the continuum ALMA observations of HD\,143006, which trace the outer disk regions, using an asymmetric Gaussian ring model, and found i\s30\dg\,and PA\s148\dg.  Their continuum dataset \citep[first presented in][]{barenfeld2016} resolves a large millimeter dust cavity ($\sim$84\,au or $\sim$0.5\as). It shows a low-contrast asymmetry in the south-east that might affect their values of $i$ and PA. 

However, the dust continuum, assuming it is optically thin, traces the disk midplane, while the scattered light signal comes from the surface layers of the disk. We therefore also consider the kinematics of the $^{12}$CO line, an optically thick tracer of the surface layers. To derive the inclination and PA of the outer disk, we consider the Moment\,1 map of the $^{12}$CO line published by \citet{barenfeld2016} and use a simple analytical model of a planar disk in Keplerian rotation around a 1.5\,M$_{\odot}$ star. We compute a projected velocity map, convolve it with a two-dimensional (2D) elliptical Gaussian beam (inferred from the ALMA data, $0.35\arcsec{}\times0.30\arcsec{}$) and fit it to the Moment\,1 map. We only consider the disk regions where the intensity is above 2$\sigma$ in the integrated intensity map. We perform our fit with the Markov chain Monte Carlo (MCMC) method  using \texttt{emcee} \citep{emcee}. Our model has five free parameters: the inclination, PA, systemic velocity, as well as right ascension and declination offsets for the center of the image. The inner radius is fixed at 2.2\,au. These parameters are sampled with uniform priors in the range of [0\dg, 90\dg], [0\dg, 360\dg],  [1 km/s, 20 km/s] and [-0.8\arcsec{}, 0.8\arcsec{}], respectively. The chain explores the parameter space with 1000 steps, with 80 walkers. We fix the outer radius of the disk to 250\,au and assume a distance of 166\,pc. We find best values for the inclination and PA of i=17.5\dg$\pm$0.2\dg\,and PA=169.8\dg$\pm$0.5\dg, and show our posterior distributions in Fig.\,\ref{fig:momCOFit}. These values are used in the deprojection of the SPHERE data to create the polar map (Fig.\,\ref{fig:obsQphiannotatedpolarmap}, right).  
We note that if the flaring of the disk is large, the thin disk model that we use is not accurate but considering the rather low inclination of the system, it is a reasonable approximation.

The discrepancy between the two estimates based on the ALMA data might be due to the two kinds of observations tracing different regions of the disk, or to the large size of the ALMA beam (\s0.35\arcsec{}, i.e., 60\,au) and the complex structure of the disk in the continuum, which makes the continuum-based estimate likely to be less reliable than the one based on  CO kinematics.

\paragraph{Inner disk.}
Inclination and PA measurements of the inner disk are very challenging due to the very high angular resolution that is required to spatially resolve the inner au. \hd was observed with the VLTI $H$-band instrument PIONIER (Precision Integrated-Optics Near-infrared Imaging ExpeRiment) in the context of a large program focused on Herbig\,AeBe stars by \citet{lazareff2017}. The $H$-band visibilities and closure phases trace the thermal emission of the hot dust located in a narrow region at the sublimation radius. In the case of HD\,143006, the $H$-band emission appears to be very compact (\s0.1\,au), and is only marginally resolved. Hence, all the analytical models considered for \hd\,(ellipsoid, ring with and without azimuthal modulation) fit the data equally well and are not well constrained. The inferred inclination and PA values are the following: i=27\dg$\pm$3\dg, PA=1\dg$\pm$13\dg\,for the ellipsoid model, respectively;  i=23\dg$\pm$5\dg, PA=168\dg$\pm$15\dg\,for the ring model with m=1 modulation; and,  i=31\dg$\pm$4\dg, PA=148\dg$\pm$21\dg\,for the ring model with m=2 modulation. Considering the very small extent of the region probed by PIONIER, and the limited angular resolution of the observations, the data can only provide a rough estimate of the inner disk geometry, as indicated by the large error bars and the strong model dependence of the results. 

The discrepancy between the various estimates of inclination and PA, as well as the features detected in the scattered light image, suggest that the inner and outer disks might be misaligned.

\paragraph{Misalignments.}
As shown by \citet{min2017}, the location and shape of the shadows seen in scattered light depend on the morphology of the inner disk, and on the shape and height of the scattering surface of the outer disk at the cavity edge, hereafter z$_{\rm{scat}}$. 
For a given orientation of the outer disk, and a given z$_{\rm{scat}}$, at the location of the shadows, the inclination and PA of the inner disk can be obtained by solving the equations that define the PA of the line connecting the shadows ($\alpha$) and the offset in declination of this line with respect to the star ($\eta$): 

$$\rm{tan}(\alpha) = \frac{\rm{sin(i_1)cos(i_2)sin(PA_1) - cos(i_1)sin(i_2)sin(PA_2)}}
{\rm{sin(i_1)cos(i_2)cos(PA_1) - cos(i_1)sin(i_2)cos(PA_2)}}$$

$$\rm{\eta} = \frac{z_{\rm{scat}} \cdot \rm{cos(i_1)}}{\rm{cos(i_2)sin(i_1)sin(PA_1) - cos(i_1)sin(i_2)sin(PA_2)}},$$

where the indices 1 and 2 refer to the inner and outer disks, respectively. For an inclined disk, at PA=0\dg, the near side of the disks is in the west. 

These equations lead to two families of solutions, depending on which side of the outer disk is the closest to us. The two possible configurations for the geometry of the inner and outer disks can  lead to similar misalignment angles. Because of the complex morphology of the object, and the fact that part of its surface is shadowed, we cannot directly infer from the scattered light observations which side of the outer disk (east or west) is the nearest to us. We therefore present the two families of solutions for the orientation of the inner disk that would reproduce the location of the shadows (in red and blue in Fig.\,\ref{fig:misalignments}). These solutions are shown as a function of z$_{\rm{scat}}/$R, with z$_{\rm{scat}}$ the height of the scattering surface of the outer disk (which differs from the pressure scale height $H_{\rm p}$ by a factor of \s2-4). The right panel shows the corresponding misalignment $\beta^{\prime}$, for the inner disk inclinations and position angles provided in the left and middle panels, respectively. As an example, assuming an outer disk inclination and PA of i=17\dg\,and PA=170\dg, a misalignment of 30\dg\,(which is the value that we will use in Sect.\,\ref{subsec:sphrt}) is obtained when the inner disk inclination and PA are \s13\dg\,and \s11\dg, respectively, and z$_{\rm{scat}}/$R is 0.12 at 0.11\arcsec{} (blue curve). If, instead, the near side of the outer disk is opposite (i.e., i=17\dg\,and PA=350\dg), such a misalignment is obtained when the inner disk inclination and PA are \s47\dg\,and \s356\dg, respectively, and z$_{\rm{scat}}/$R = 0.17 at the outer disk rim location (red curve). Assuming that the scattering surface corresponds to \s2-4 $H_{\rm p}$, we find that the disk aspect ratio $H_{\rm p}$/R is \s0.03-0.08 at the outer disk rim.    

%

\subsection{Hydrodynamic \& radiative transfer model}
\label{subsec:sphrt}
In earlier studies, a parametric approach  was used to determine the disk geometry and density structure in the inner and outer disks that would lead to the observed shadowing pattern seen in scattered light observations of protoplanetary disks \citep[e.g.,][]{marino2015,benisty2017}.  To model HD\,143006, we use 3D hydrodynamical simulations, first presented in \citet{facchini2018}. Our observations (Fig.~\ref{fig:obsQphi}) are strikingly similar to their predictions  (see their right panel of Fig.~10), in particular regarding the east/west brightness asymmetry. 

We provide here a summary of the simulations, and for more details, we refer the reader to \citet{facchini2018}. The 3D simulations have been performed with the Smoothed Particle Hydrodynamics (SPH) code \textsc{phantom} \citep{2017arXiv170203930P}, using $10^6$ particles. We consider a protoplanetary disk and an equal mass binary with a semi-major axis $a_0$, and inclined by $60^\circ$ with respect to the disk. The disk has an initial surface density scaling with $r^{-1}$, with $r$ being the radial coordinate. The temperature profile in the simulation is taken to be vertically isothermal, with the temperature scaling as $T\propto r^{-1/2}$, and an aspect ratio of $H_{\rm p}/r=0.041$ at $r=1.7a_0$.  After a few binary orbits the circumbinary disk breaks into two separate annuli, driven by the tidal torques generated by the binary on an inclined orbit. Once the inner disk disconnects from the outer disk, it precesses freely around the binary angular momentum vector \citep{2013MNRAS.433.2157L}. The inclined ring extends from 1.7$a_0$ to 5$a_0$. For this specific setup, the inner and outer disk can show a mutual misalignment between $10^\circ$ and $110^\circ$, with the angle varying as the inner disk precesses. 

To generate synthetic observables from the SPH hydrodynamic simulations, we assume that the small dust grains, which scatter light efficiently, and the gas, are dynamically coupled, and  use the 3D radiative transfer code \textsc{radmc-3d}\footnote{\url{http://www.ita.uni-heidelberg.de/~dullemond/software/radmc-3d/}}. We note that the temperature profiles of the hydrodynamical and radiative transfer simulations are not computed self-consistently. We first  scale the hydrodynamic simulations such that the binary orbital separation $a_0$ is 5.2\,au.  The disk aspect ratio being $H_{\rm p}/r=0.041$ at 8.8\,au, with a flaring index of 0.25, implies that at the outer disk location (0.11\arcsec{}, i.e. \s18.3\,au), $H_{\rm p}/r=0.05$, consistent with the estimate obtained from the relative misalignment of the disks and the location of the shadows (see Sect.\,\ref{sec:almaimaging}).   After the scaling of the simulations, we interpolate the particle-based density distribution in the SPH simulations to the 3D spherical mesh used in the radiative transfer calculations, using the standard cubic spline kernel. The spherical mesh consists of $N_{\rm r} = 220$, $N_{\theta} = 200$, and $N_{\phi} = 200$ grid cells in the radial, poloidal and azimuthal directions in the  intervals [7.8\,au, 83\,au], [0, $\pi$], [0, 2$\pi$]. 

To compare to our SPHERE images, we post-process a snapshot corresponding to 245 binary orbits, for which the misalignment between the inner and outer disk is \s$30^\circ$ (see Fig.\,\ref{fig:sph}). In this snapshot, the outer disk inclination is \s16\dg, and its PA is \s170\dg, while for the inner disk the inclination and PA are \s16\dg\,and 14\dg, respectively. As explained in Sect.\,\ref{sec:almaimaging}, since we do not know which side of the outer disk is closer to us, we consider a second solution with \s14\dg\,and \s350\dg\,as the inclination and PA of the outer disk, respectively, and \s44\dg\,and \s355\dg\,for the inclination and PA of the inner disk.  In the first solution, while the inclination with respect to the line of sight is the same, the orientation of the two disks is almost opposite: the east side of the outer disk is closer to us, while the near side of the inner disk is in the west. In the second solution, the near side of both the inner and outer disk is the wwestest side, but the inclinations differ by 30\dg. These values are close to the estimates derived in Sect.\,\ref{sec:almaimaging}, based on the location of the shadows.  We note however, that these values are model-dependent as they depend on the density considered in the inner disk and within the gap separating it from the outer disk. We therefore expect that other types of models (e.g. with a small circumprimary disk instead of a circumbinary disk) would provide slightly different values as long as the misalignment is moderate (\s20-30\dg). 



\begin{figure*}
\includegraphics[width=1.15\textwidth]{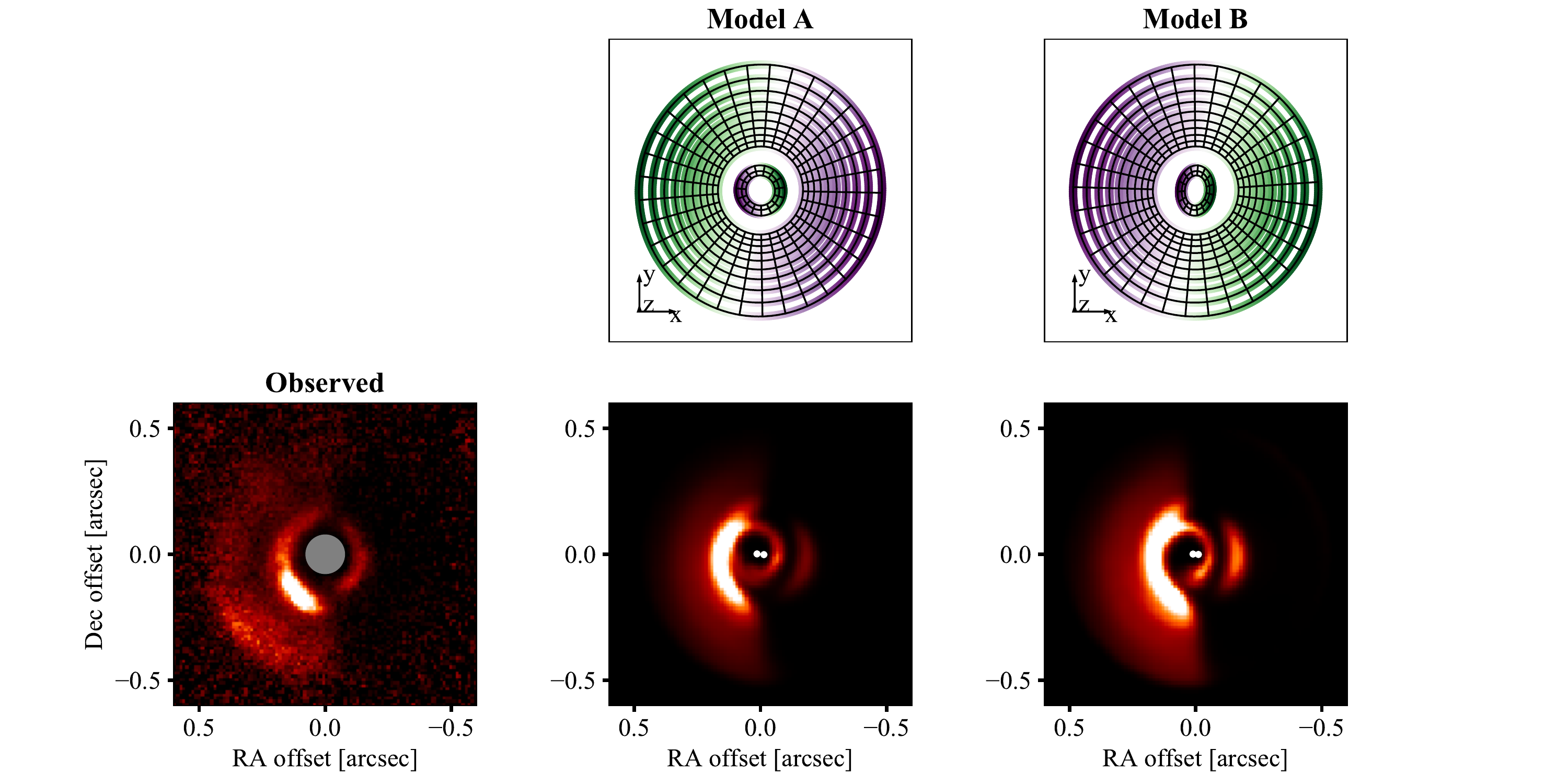}
\caption{Scattered light predictions at 1.2\,$\mu$m of the hydrodynamic model (bottom middle and bottom right panels) with a misalignment angle between the inner and outer disk of $\sim$30\dg, compared to the observations (bottom left panel). The locations of the two stars are indicated with white circles. All images are scaled by r$^{2}$. The upper panels show a schematic of the 3D structure, with a color coding (green, purple) that indicates the regions of the disk above and below the plane perpendicular to the line of sight. Model A corresponds to the blue solution of Fig.\,\ref{fig:misalignments}, model B to the red.}
\label{fig:rtmodel}
\end{figure*}

We use the stellar parameters mentioned in the Introduction. \citet{barenfeld2016} derived a dust mass of \s24.3 Earth masses for the disk, by converting the sub-millimeter continuum flux using d=145\,pc. We scale this value using the new Gaia distance (d=166\,pc), and consider $9.5\cdot10^{-3}$M$_\odot$. 
To be consistent with the hydrodynamical simulations, we used two identical stars that reproduce the total stellar luminosity of HD\,143006.  The dust opacity and scattering matrix elements were calculated from the optical constants of astronomical silicates \citep{weingartner_2001} for a grain size distribution of $n(a)\propto a^{-3.5}$ for grain sizes between $a=0.1$\,microns and $a=1$\,mm.  We use  $10^8$ photon packages to calculate the dust temperature in a thermal Monte Carlo simulation as well as to calculate scattered light images in the $J$ band (1.2\,$\mu$m).  Once the synthetic images are computed, we convolve the scattered light predictions with a FLUX image from the dataset, to reduce their resolution to that of the observations. 

The model predictions are shown in Fig.\,\ref{fig:rtmodel}. The inner disk is inclined and casts a shadow onto the outer disk at two points (similarly to the cases of HD\,142527 and HD\,100453), but as the inclination is only moderate, the shadow also darkens half of the outer disk. Because we do not know which side of the disk is closer to us, we provide two solutions in  Fig.\,\ref{fig:rtmodel}  (model A and B). The upper panels show a sketch of the 3D structure of the disk, with a color coding that indicates which part of the disk is above or below the plane perpendicular to our line of sight.  Our model reproduces most of the features observed in the scattered light observations: a clear east/west brightness asymmetry, two narrow shadows,  and two bright arcs tracing Ring \#2. We note that the circumbinary disk clearly appears in the synthetic image, while it is not detected in the observations up 60\,mas (i.e., \s10\,au). This supports the presence of a small, misaligned circumprimary disk rather than a circumbinary disk. This is discussed further in Sect.\,\ref{sec:discussion}. We note that the model, and the consequent shadowing due to a misaligned inner disk, cannot reproduce the bright region observed along PA\s110-170\dg, in particular along Ring\,\#2, nor the outermost gap (Gap\,\#1). 


\section{Discussion}
\label{sec:discussion}
Warps have been inferred in many protoplanetary disks, with various observational tracers. Shadows in scattered light appear as steady low brightness regions \citep{stolker2016,benisty2017,casassus2018}, for which a moderate to large misalignment between the inner and outer disk up to \s70\dg\,was suggested from radiative transfer modeling. In one of them, HD\,142527, a stellar companion on an eccentric orbit is thought to be responsible for the misalignment \citep{price2018}. In other objects, such as SAO\,206462 and RXJ1604.3-2130A, the shadows appear very variable  in amplitude, width,  and location \citep[][Pinilla et al. submitted]{stolker2016b}. For example, in RXJ1604.3-2130A, an object known to be an aperiodic dipper \citep{ansdell2016}, we find that the timescale for the variations is shorter than a day, indicating a very complex and dynamic inner disk (Pinilla et al. submitted). In AA\,Tau, a strong, inclined magnetic field induces a warp at the disk inner edge that periodically rotates with the stellar period \citep{bouvier2007}. However, recent observations by \citet{loomis2017} indicate that the inner disk is also perturbed, with an additional warp and evidence for a radial inflow, possibly due to gap-crossing streamers. Another case of a perturbed inner disk with a warp is V354\,Mon, which presents a low gas-to-dust ratio in the inner disk, and dimming events that could be due to small dust particles that results from the fragmentation of larger particles that drift from the outer disk \citep{schneider2018}. In all these objects, the presence of a companion in the stellar or sub-stellar mass regime, at a separation of a couple of tens of \,au, could explain some characteristics of the observations.

\subsection{Origin of the warp}
In this paper, we consider a model of a broken and misaligned circumbinary disk due to the gravitational influence of a stellar companion on an inclined orbit. It successfully reproduces the general  characteristics of the scattered light image, supporting the idea that the disk of HD\,143006 hosts a warp inside the observed cavity.  We stress that our observational predictions hold for any warped disk with a moderate misalignment, independently of what is causing the torque that leads to the misalignment. In particular, as HD\,143006 shows a near infrared excess, indicating the presence of hot dust grains very close to the star, it is likely that it hosts a circumprimary disk that is possibly tilted, rather than a circumbinary disk, as in the HD\,142527 system. However, since we do not have any direct image of the innermost regions, we cannot directly determine the outer extent of the inner disk, the location of the warp, or the location of the putative companion inducing it.

\paragraph{Upper limits on the binarity.} \citet{kraus2008} led a survey on the stellar binarity of 82 young stars from  Upper Sco, using non-redundant aperture masking interferometry, a technique that allows us to search for companions at the diffraction limits.  Combining their results with the ones from the literature, they report a frequency of binary companions of \s33$^{+5}_{-4}$\% at separations of 6 to 435\,au, using d=145\,pc as the distance of Upper Sco. For HD\,143006, they estimate lower limits on the $K$-band contrast of \s3.5 for separations within 20 to 40\,mas, and of \s5.1 within 40-80\,mas, that is, detection limits of companions with $K$-band apparent magnitude of 10.6 and 12.2, respectively. Using the BT-SETTL models \citep{baraffe2015}, considering d=166\,pc, and assuming an age of 10\,Myrs, these detection limits translate into companion masses of 0.45\,M$_{\odot}$ and 0.16\,M$_{\odot}$, corresponding to a mass ratio of q=0.3 and 0.1, respectively. 
Using archival near-infrared  ($H$-band)  interferometric (VLTI/ PIONIER) observations of  \hd, we performed an analysis of the closure phase data, assuming that the inner disk is point-symmetric and that any departure from point-symmetry detected in the closure phase would be due to the presence of a binary companion. We used the Companion Analysis and Non-Detection in Interferometric Data algorithm \citep[CANDID,][]{gallenne2015}, which allows an estimate of the 3-$\sigma$ detection limit for any companion at varying separation from the primary star. Similarly to \citet{kraus2008}, we find that a companion with a $H$-band contrast lower than 3.7 magnitude would have been detected at more than 3-$\sigma$ within 150\,mas (i.e., 25\,au, bandwidth smearing limitation), which translates into a mass ratio of q= 0.2 using the parameters mentioned above. 
%
%
\paragraph{A warp induced by a companion.}
Considering these detection limits, the presence of an equal mass binary can be excluded, but a low mass stellar object, or a massive planet, could still be present in the cavity of HD\,143006.  For example, assuming that \hd has a stellar mass of 1.5\,M$_{\odot}$, a 10 M$_{\rm{Jup}}$  planet  (q=0.006) would not be detected by the interferometric observations.  
Such a massive object  could possibly be responsible for the misalignment of a circumprimary disk.  If the inner disk angular momentum is lower than that of an inclined planet, the latter can tilt the disk inside its orbital radius  \citep[e.g.,][]{matsakos2017}.   If the planet is massive enough to carve a gap, the inner and outer disk separate and can be both tilted with respect to the midplane \citep[e.g.,][]{bitsch2013,nealon2018}. For a 6-M$_{\rm{Jup}}$ planet with initial inclinations ranging from 10 to 80\dg, the misalignment can be up to 15\dg\, \citep{xg2013}. 

\citet{owen2017} recently investigated a mechanism that can lead to large misalignments with a companion-star system that is originally coplanar.  They find that, for a stellar/companion mass ratio of 0.01-0.1 and separation of 10-100\,au, secular precession resonances can generate large misalignments between the inner (circumprimary) and outer (circumbinary) disks. The resonance between the inner disk and the companion can lead to a wide range of misalignment angles between the inner and outer disk, some as large as the ones observed for HD\,142527 and HD\,100453, while the companion remains in the plane of the outer disk. Finally, a massive planet could become misaligned via the  Kozai effect \citep{martin2016}, when an additional external companion is present. However, this scenario does not seem plausible, since there is no hint of a binary companion orbiting outside the circumstellar disk.


In general, it is unclear if the disk would be broken and misaligned at a specific stage of its evolution. In the conditions of our hydrodynamical simulations, disk breaking is favored for a low disk aspect ratio, therefore a misaligned companion could more easily break the disk when the star gets older and colder (which in turn implies a colder disk and lower aspect ratio). This would support the findings of Garufi et al. (submitted), that shadows observed in scattered light, smoking guns of misalignments, are primarily found around rather old objects. Interestingly, these are often also the objects with higher photospheric metallicity \citep{banzatti2018}, showing spiral arms, and with the highest near infrared excess, like in the case of HD\,143006 (\s21\%), which indicates a strongly inflated inner disk with micron-sized dust grains reprocessing stellar light at high altitude above the disk midplane. This might be a result of the interaction of the inner disk with an inclined companion. 
If such a companion creates a gap or cavity, which is depleted with time, the inner disk will lose mass without efficient replenishment from the outer disk, and at some point the angular momentum of the companion will become larger than the one from the inner disk, favoring a late misalignment between the inner and outer disk.

The misalignment of the companion(s) could also occur very early in a disk's history and be an imprint from the early stages of star formation. Misaligned planets could form by fragmentation in the accretion phase of the protostellar envelope \citep{terquem2002} and while most are ejected from the system, some could remain in an inclined orbit around the star. In this framework, \citet{teyssandier2013} find that massive planets  (approximately the mass of Jupiter) would be circularized and align in the disk midplane but that Neptune-mass planets would remain on inclined orbits over the disk lifetime.  They note however, that if the disk mass steeply decreases with time, more massive objects could remain misaligned. Interestingly, a Neptune-mass planet would allow a continuous replenishment of small dust in the inner disk from the outer disk \citep{pinilla2015,rosotti2016}, which is needed in the case of HD\,143006.


\paragraph{Other scenarios.}
Apart from being induced by a misaligned planet, inner disk warps can be due to a strong misaligned dipolar magnetic field as in the well-documented case of AA\,Tau. \citet{lavail2017} find a magnetic field of 1.4 kG in \hd (called V1149 Sco in their paper), but the data are not sufficient to obtain a topology of the magnetic field. If the magnetic field is strongly inclined, and warps the inner disk edge, such a warp would rotate with the stellar period (approximately a few days), and lead to fast changing shadows. This scenario can easily be tested with multi-epoch observations. Unfortunately, our two $J$-band epochs (with and without coronagraph) were obtained on the same observing night but a low signal to noise ratio optical image, obtained a year before with the ZIMPOL (Zurich IMaging POLarimeter) instrument is shown in the appendix Fig.\,\ref{fig:zim}. Although the low quality of the image prevents a detailed analysis of the disk image, we find that the brightest features (east/west sides of Ring\,\#2) have not moved significantly, and hence that it is likely that the shadows do not move at the stellar period. 

Another scenario that would lead to misalignments of inner and outer disk regions, is if they are primordial ones, due to a late accretion of material that has an angular momentum misaligned with that of the star (Bate et al. 2010, Dullemond et al. submitted). Recent hydrodynamical simulations carried out by \citet{bate2018} lead to such a case, with a circumbinary disk whose inner and outer disk planes differ, but this is not a common outcome. Instead, many of the  simulations with multiple stars lead to circumstellar and/or circumbinary disks that can be misaligned with each other  \citep{bate2018}. 


\subsection{Rings, gaps, and brightness asymmetries}
Our scattered light image shows two rings and two gaps. The outer ring (Ring\,\#1) might be tracing the outer disk up to what our sensitivity allows and might therefore not be tracing any ring-like perturbation (in density or scale height) of the disk. Gap\,\#1 shows significant scattered light signal, and does not appear empty of small grains, in particular in the region located between PAs 110\dg\,and 170\dg, as seen in Figs.\,\ref{fig:obsQphiannotatedpolarmap} and \ref{fig:radazicuts}, right, in which the disk appears almost continuous. Gap\,\#1 could be tracing a marginal depletion in small dust, in a gap opened by a planet \citep[e.g.,][]{dong2017}, while Ring\,\#2 would be its inner edge. It is also possible that Gap\,\#1 is due to self-shadowing by the inner edge of the outer disk (Ring\,\#2).  In that case, the denomination of Ring \#1 is artificial, in the sense that it would trace the outer disk illumination beyond the shadow of Ring\,\#2 rather than a depleted region.  By contrast, Gap\,\#2 appears to be quite depleted in dust, so Ring\,\#2 could be directly irradiated by the star and puffed up. This would naturally lead to Ring \#2 casting a shadow on the outer disk. 

Modeling of the visibility data of ALMA observations reveals that there is a large cavity in the millimeter emission, surrounded by an asymmetric ring peaking at 84\,au \citep[0.5\as,][]{pinilla2018b}. This location coincides with the outer edge of Ring\,\#1 that extends from \s0.3\as to \s0.5\as\,and peaks approximately at \s0.45\as. If this outer ring is a real density enhancement (instead of being only an effect of self-shadowing), Ring\,\#1 and the asymmetric ALMA ring may have a common origin due to dust trapping in pressure bumps possibly induced by a planetary companion. 
On the other hand, if Ring\,\#1 is not a density enhancement and if it only results from self-shadowing, the very large radial segregation between the inner ring (Ring\,\#2) and the ring observed with ALMA cannot be explained by a single giant planet, and instead, a stellar-mass companion would be required. 

The near-infrared excess indicates the existence of an optically thick dusty belt located close to the dust sublimation radius within the first few astronomical units. At these distances from the star, the gas density is expected to be high enough that the millimeter dust particles too should be coupled to the gas. Therefore, with very high angular resolution ($\sim$0.01-0.02\as) observations with ALMA, it may be possible to detect a misaligned inner disk in the millimeter emission as well, and confirm our findings based on the shadows observed in the scattered light image. 

Even though a misaligned disk model is successful in reproducing the east/west asymmetry, it does not account for the narrow overbrightness between PA 110\dg\,and 170\dg\,(see Fig.\,\ref{fig:rtmodel}). Azimuthal asymmetries in scattered light can be due to the scattering angle and polarization efficiency of dust grains located on the disk surface. Large grains ($\geq5\mu$m) are efficient forward scatterers, often leading to one side of the disk (the near-side) being brighter than the other. Assuming that the PA of the outer disk is \s170\dg, we would expect to see the overall east or west side brighter than the other, and not over such an azimuthally narrow region. The polarization efficiency being maximum for 90\dg\, scattering angles can in turn lead to bright lobes along the major axis of the disk for inclined disks \citep[see, e.g., the radiative transfer model in ][]{benisty2017}. The disk of HD\,143006 being only very slightly inclined, the latter possibility cannot account for the strong brightness asymmetry.  It is possible that the disk presents a local overdensity, for example, due to the formation of a vortex, an eccentric ring, or a spiral arm at the outer edge of the cavity. If the overbrightness traces the tip of a spiral arm, we would expect it to not be co-radial, as observed for Ring\,\#2.  Interestingly, the continuum ALMA data \citep{barenfeld2016, pinilla2018b} show a low contrast asymmetry (contrast less than a factor of two) along the same angle as the one observed in the scattered light data, supporting a density enhancement. However, it is located close to the outer ring (Ring\,\#1), at  large radii (in the ALMA image, peaking at 0.5\as or 84\,au), and does not coincide with the inner ring (Ring\,\#2). From the current ALMA observations, the asymmetry is not resolved and its exact morphology is still an open question.   As shown in the polar map, not only Ring\,\#2 shows the overbrightness, but the outer disk does too. This suggests that along this  range in PAs, the disk is more strongly illuminated than the rest of the disk. An equally possible scenario is that most of the disk (between PAs of 0 and 110\dg) lies in a partial shadow caused by the inner disk, while the bright region (PAs 110-170\dg) is unshadowed. This additional shadow could be due to asymmetric features from the inner disk that have a moderate radial optical depth, likely from tenuous surface layers. In either scenario, the physical cause of this phenomenon is  not clear.

\section{Conclusions}
In this paper, we present the first scattered light observations of the circumstellar disk around the T Tauri star HD\,143006. Our observations reveal two rings and two gaps, a strong east/west brightness asymmetry, an  overbrightness along a narrow range of position angles (110-170\dg), and two dark narrow lanes. Such azimuthal brightness variations are indicative of shadowing effects, in particular due to a misaligned inner disk. 

We analyze the kinematics of the $^{12}$CO line, as observed with ALMA, using an analytical model of a razor-thin disk in Keplerian rotation and derive an inclination of \s17\dg\,and a position angle of 170\dg\,for the outer disk. Combined with  the constraints derived from inner disk observations with near infrared interferometry, this suggests that the inner and outer disk regions are moderately misaligned. We provide two possible solutions for the inner disk orientation to reproduce the location of the shadows, depending on the side of the outer disk that is nearest to us. 

Our scattered light image shares a striking resemblance with synthetic predictions based on hydrodynamical simulations of a protoplanetary disk warped by an inclined equal mass binary \citep{facchini2018}. In these simulations, the circumbinary disk breaks into two distinct annuli (inner and outer disk, both circumbinary), and the inner disk precesses freely around the binary angular momentum vector. To compare with our observations, we post-process a snapshot of these simulations for which the relative misalignment between the two annuli is \s30\dg, considering the stellar parameters of HD\,143006. This model reproduces the east/west asymmetry, but does not account for the additional overbrightness along a narrow range of position angles which might be due to an overdensity not included in our model. 

Although our model uses an equal mass binary, which can be ruled out by current detection limits, we stress that our predictions hold for any warped disk, independently of the cause of the misalignment. In particular, a massive planet (e.g. with a mass ratio of 0.01-0.1) might break the disk, or alternatively, an inclined magnetic field  could misalign the innermost disk edge as in AA\,Tau, although our marginal evidence that the shadows have not rotated within a year does not support the latter scenario. 

Further observations of this system, in particular with ALMA at high resolution, will allow us to constrain the orientation of the inner disk in large (mm) grains, and could confirm or contradict that the features observed in our scattered light images are due to a misaligned inner disk. Comparison between  ALMA and SPHERE observations at similar angular resolution will also allow us to constrain whether the disk shows evidence for cooler regions in the mm, due to the shadows, as in DoAr\,44 \citep{casassus2018}.

Scattered light shadows have now been found in a handful of objects, often in transition disks with large cavities that could host a high  planetary mass or a low stellar mass companion \citep[as in HD\,142527,][]{biller2012}, which would still be below the current detection limits. It is therefore possible that all transition disks host stellar or planetary-mass companions with a mass ratio of \s0.01, some with inclined orbits. It is unclear however, if the misalignments that the shadows observed in scattered light trace could be the origin of the relative inclinations between the stellar rotation axis and orbit orientation found in many exoplanetary systems.


\section*{Acknowledgements}
We thank the referee, Ruobing Dong, for constructive comments that helped improve the manuscript.  We are thankful to A.\,Triaud, Z.\, Zhu, R.\,Nealon, C.\,Manara, and J.\,Huang  for insightful  discussions. We thank S.\,Barenfeld for sharing his ALMA data of HD\,143006.  M.B. and M.V. acknowledge funding from ANR of France under contract number ANR-16-CE31-0013 (Planet Forming disks). This work has been supported by the diskSIM project, grant agreement 341137 funded by the European Research Council under ERC-2013-ADG.  P.P. acknowledges support by NASA through Hubble Fellowship grant HST-HF2-51380.001-A awarded by the Space Telescope Science Institute, which is operated by the Association of Universities for Research in Astronomy, Inc., for NASA, under contract NAS 5-26555. The  figures were generated with the \textsc{python}-based package \textsc{matplotlib} \citep{matplotlib}.

\bibliographystyle{aa}
\bibliography{hd143006_vap}

\begin{thebibliography}{96}
\expandafter\ifx\csname natexlab\endcsname\relax\def\natexlab#1{#1}\fi

\bibitem[{{ALMA Partnership} {et~al.}(2015){ALMA Partnership}, {Brogan},
  {P{\'e}rez}, {Hunter}, {Dent}, {Hales}, {Hills}, {Corder}, {Fomalont},
  {Vlahakis}, {Asaki}, {Barkats}, {Hirota}, {Hodge}, {Impellizzeri}, {Kneissl},
  {Liuzzo}, {Lucas}, {Marcelino}, {Matsushita}, {Nakanishi}, {Phillips},
  {Richards}, {Toledo}, {Aladro}, {Broguiere}, {Cortes}, {Cortes}, {Espada},
  {Galarza}, {Garcia-Appadoo}, {Guzman-Ramirez}, {Humphreys}, {Jung}, {Kameno},
  {Laing}, {Leon}, {Marconi}, {Mignano}, {Nikolic}, {Nyman}, {Radiszcz},
  {Remijan}, {Rod{\'o}n}, {Sawada}, {Takahashi}, {Tilanus}, {Vila Vilaro},
  {Watson}, {Wiklind}, {Akiyama}, {Chapillon}, {de Gregorio-Monsalvo}, {Di
  Francesco}, {Gueth}, {Kawamura}, {Lee}, {Nguyen Luong}, {Mangum}, {Pietu},
  {Sanhueza}, {Saigo}, {Takakuwa}, {Ubach}, {van Kempen}, {Wootten},
  {Castro-Carrizo}, {Francke}, {Gallardo}, {Garcia}, {Gonzalez}, {Hill},
  {Kaminski}, {Kurono}, {Liu}, {Lopez}, {Morales}, {Plarre}, {Schieven},
  {Testi}, {Videla}, {Villard}, {Andreani}, {Hibbard}, \&
  {Tatematsu}}]{2015ApJ...808L...3A}
{ALMA Partnership}, {Brogan}, C.~L., {P{\'e}rez}, L.~M., {et~al.} 2015, \apjl,
  808, L3

\bibitem[{{Andrews} {et~al.}(2016){Andrews}, {Wilner}, {Zhu}, {Birnstiel},
  {Carpenter}, {P{\'e}rez}, {Bai}, {{\"O}berg}, {Hughes}, {Isella}, \&
  {Ricci}}]{andrews2016}
{Andrews}, S.~M., {Wilner}, D.~J., {Zhu}, Z., {et~al.} 2016, \apjl, 820, L40

\bibitem[{{Ansdell} {et~al.}(2016){Ansdell}, {Gaidos}, {Williams}, {Kennedy},
  {Wyatt}, {LaCourse}, {Jacobs}, \& {Mann}}]{ansdell2016}
{Ansdell}, M., {Gaidos}, E., {Williams}, J.~P., {et~al.} 2016, \mnras, 462,
  L101

\bibitem[{{Apai} {et~al.}(2004){Apai}, {Pascucci}, {Brandner}, {Henning},
  {Lenzen}, {Potter}, {Lagrange}, \& {Rousset}}]{apai2004}
{Apai}, D., {Pascucci}, I., {Brandner}, W., {et~al.} 2004, \aap, 415, 671

\bibitem[{{Avenhaus} {et~al.}(2018){Avenhaus}, {Quanz}, {Garufi}, {Perez},
  {Casassus}, {Pinte}, {Bertrang}, {Caceres}, {Benisty}, \&
  {Dominik}}]{avenhaus2018}
{Avenhaus}, H., {Quanz}, S.~P., {Garufi}, A., {et~al.} 2018, \apj, 863, 44

\bibitem[{{Avenhaus} {et~al.}(2014){Avenhaus}, {Quanz}, {Schmid}, {Meyer},
  {Garufi}, {Wolf}, \& {Dominik}}]{avenhaus2014}
{Avenhaus}, H., {Quanz}, S.~P., {Schmid}, H.~M., {et~al.} 2014, \apj, 781, 87

\bibitem[{{Banzatti} {et~al.}(2018){Banzatti}, {Garufi}, {Kama}, {Benisty},
  {Brittain}, {Pontoppidan}, \& {Rayner}}]{banzatti2018}
{Banzatti}, A., {Garufi}, A., {Kama}, M., {et~al.} 2018, \aap, 609, L2

\bibitem[{{Baraffe} {et~al.}(2015){Baraffe}, {Homeier}, {Allard}, \&
  {Chabrier}}]{baraffe2015}
{Baraffe}, I., {Homeier}, D., {Allard}, F., \& {Chabrier}, G. 2015, \aap, 577,
  A42

\bibitem[{{Barenfeld} {et~al.}(2016){Barenfeld}, {Carpenter}, {Ricci}, \&
  {Isella}}]{barenfeld2016}
{Barenfeld}, S.~A., {Carpenter}, J.~M., {Ricci}, L., \& {Isella}, A. 2016,
  \apj, 827, 142

\bibitem[{{Bate}(2018)}]{bate2018}
{Bate}, M.~R. 2018, \mnras, 475, 5618

\bibitem[{{Benisty} {et~al.}(2015){Benisty}, {Juhasz}, {Boccaletti},
  {Avenhaus}, {Milli}, {Thalmann}, {Dominik}, {Pinilla}, {Buenzli}, {Pohl},
  {Beuzit}, {Birnstiel}, {de Boer}, {Bonnefoy}, {Chauvin}, {Christiaens},
  {Garufi}, {Grady}, {Henning}, {Huelamo}, {Isella}, {Langlois}, {M{\'e}nard},
  {Mouillet}, {Olofsson}, {Pantin}, {Pinte}, \& {Pueyo}}]{benisty2015}
{Benisty}, M., {Juhasz}, A., {Boccaletti}, A., {et~al.} 2015, \aap, 578, L6

\bibitem[{{Benisty} {et~al.}(2017){Benisty}, {Stolker}, {Pohl}, {de Boer},
  {Lesur}, {Dominik}, {Dullemond}, {Langlois}, {Min}, {Wagner}, {Henning},
  {Juhasz}, {Pinilla}, {Facchini}, {Apai}, {van Boekel}, {Garufi}, {Ginski},
  {M{\'e}nard}, {Pinte}, {Quanz}, {Zurlo}, {Boccaletti}, {Bonnefoy}, {Beuzit},
  {Chauvin}, {Cudel}, {Desidera}, {Feldt}, {Fontanive}, {Gratton}, {Kasper},
  {Lagrange}, {LeCoroller}, {Mouillet}, {Mesa}, {Sissa}, {Vigan}, {Antichi},
  {Buey}, {Fusco}, {Gisler}, {Llored}, {Magnard}, {Moeller-Nilsson}, {Pragt},
  {Roelfsema}, {Sauvage}, \& {Wildi}}]{benisty2017}
{Benisty}, M., {Stolker}, T., {Pohl}, A., {et~al.} 2017, \aap, 597, A42

\bibitem[{{B{\'e}thune} {et~al.}(2016){B{\'e}thune}, {Lesur}, \&
  {Ferreira}}]{bethune2016}
{B{\'e}thune}, W., {Lesur}, G., \& {Ferreira}, J. 2016, \aap, 589, A87

\bibitem[{{Beuzit} {et~al.}(2008){Beuzit}, {Feldt}, {Dohlen}, {Mouillet},
  {Puget}, {Wildi}, {Abe}, {Antichi}, {Baruffolo}, {Baudoz}, {Boccaletti},
  {Carbillet}, {Charton}, {Claudi}, {Downing}, {Fabron}, {Feautrier},
  {Fedrigo}, {Fusco}, {Gach}, {Gratton}, {Henning}, {Hubin}, {Joos}, {Kasper},
  {Langlois}, {Lenzen}, {Moutou}, {Pavlov}, {Petit}, {Pragt}, {Rabou}, {Rigal},
  {Roelfsema}, {Rousset}, {Saisse}, {Schmid}, {Stadler}, {Thalmann}, {Turatto},
  {Udry}, {Vakili}, \& {Waters}}]{beuzit2008}
{Beuzit}, J.-L., {Feldt}, M., {Dohlen}, K., {et~al.} 2008, in SPIE Proc., Vol.
  7014, 18

\bibitem[{{Biller} {et~al.}(2012){Biller}, {Lacour}, {Juh{\'a}sz}, {Benisty},
  {Chauvin}, {Olofsson}, {Pott}, {M{\"u}ller}, {Sicilia-Aguilar}, {Bonnefoy},
  {Tuthill}, {Thebault}, {Henning}, \& {Crida}}]{biller2012}
{Biller}, B., {Lacour}, S., {Juh{\'a}sz}, A., {et~al.} 2012, \apjl, 753, L38

\bibitem[{{Bitsch} {et~al.}(2013){Bitsch}, {Crida}, {Libert}, \&
  {Lega}}]{bitsch2013}
{Bitsch}, B., {Crida}, A., {Libert}, A.~S., \& {Lega}, E. 2013, \aap, 555, A124

\bibitem[{{Bodman} {et~al.}(2017){Bodman}, {Quillen}, {Ansdell}, {Hippke},
  {Boyajian}, {Mamajek}, {Blackman}, {Rizzuto}, \& {Kastner}}]{bodman2017}
{Bodman}, E. H.~L., {Quillen}, A.~C., {Ansdell}, M., {et~al.} 2017, \mnras,
  470, 202

\bibitem[{{Boehler} {et~al.}(2018){Boehler}, {Ricci}, {Weaver}, {Isella},
  {Benisty}, {Carpenter}, {Grady}, {Shen}, {Tang}, \& {Perez}}]{boehler2018}
{Boehler}, Y., {Ricci}, L., {Weaver}, E., {et~al.} 2018, \apj, 853, 162

\bibitem[{{Bouvier} {et~al.}(2007){Bouvier}, {Alencar}, {Boutelier},
  {Dougados}, {Balog}, {Grankin}, {Hodgkin}, {Ibrahimov}, {Kun}, {Magakian}, \&
  {Pinte}}]{bouvier2007}
{Bouvier}, J., {Alencar}, S.~H.~P., {Boutelier}, T., {et~al.} 2007, \aap, 463,
  1017

\bibitem[{{Brinch} {et~al.}(2016){Brinch}, {J{\o}rgensen}, {Hogerheijde},
  {Nelson}, \& {Gressel}}]{brinch2016}
{Brinch}, C., {J{\o}rgensen}, J.~K., {Hogerheijde}, M.~R., {Nelson}, R.~P., \&
  {Gressel}, O. 2016, \apj, 830, L16

\bibitem[{{Canovas} {et~al.}(2011){Canovas}, {Rodenhuis}, {Jeffers}, {Min}, \&
  {Keller}}]{canovas2011}
{Canovas}, H., {Rodenhuis}, M., {Jeffers}, S.~V., {Min}, M., \& {Keller}, C.~U.
  2011, \aap, 531, A102

\bibitem[{{Carbillet} {et~al.}(2011){Carbillet}, {Bendjoya}, {Abe}, {Guerri},
  {Boccaletti}, {Daban}, {Dohlen}, {Ferrari}, {Robbe-Dubois}, {Douet}, \&
  {Vakili}}]{carbillet2011}
{Carbillet}, M., {Bendjoya}, P., {Abe}, L., {et~al.} 2011, Experimental
  Astronomy, 30, 39

\bibitem[{{Casassus} {et~al.}(2018){Casassus}, {Avenhaus}, {P{\'e}rez},
  {Navarro}, {C{\'a}rcamo}, {Marino}, {Cieza}, {Quanz}, {Alarc{\'o}n}, {Zurlo},
  {Osses}, {Rannou}, {Rom{\'a}n}, \& {Barraza}}]{casassus2018}
{Casassus}, S., {Avenhaus}, H., {P{\'e}rez}, S., {et~al.} 2018, \mnras, 868

\bibitem[{{Casassus} {et~al.}(2015){Casassus}, {Marino}, {P{\'e}rez}, {Roman},
  {Dunhill}, {Armitage}, {Cuadra}, {Wootten}, {van der Plas}, {Cieza}, {Moral},
  {Christiaens}, \& {Montesinos}}]{casassus2015}
{Casassus}, S., {Marino}, S., {P{\'e}rez}, S., {et~al.} 2015, \apj, 811, 92

\bibitem[{{Casassus} {et~al.}(2013){Casassus}, {van der Plas}, {M}, {Dent},
  {Fomalont}, {Hagelberg}, {Hales}, {Jord{\'a}n}, {Mawet}, {M{\'e}nard},
  {Wootten}, {Wilner}, {Hughes}, {Schreiber}, {Girard}, {Ercolano}, {Canovas},
  {Rom{\'a}n}, \& {Salinas}}]{2013Natur.493..191C}
{Casassus}, S., {van der Plas}, G., {M}, S.~P., {et~al.} 2013, \nat, 493, 191

\bibitem[{{Cieza} {et~al.}(2017){Cieza}, {Casassus}, {P{\'e}rez}, {Hales},
  {C{\'a}rcamo}, {Ansdell}, {Avenhaus}, {Bayo}, {Bertrang}, {C{\'a}novas},
  {Christiaens}, {Dent}, {Ferrero}, {Gamen}, {Olofsson}, {Orcajo}, {Osses},
  {Pe{\~n}a-Ramirez}, {Principe}, {Ru{\'{\i}}z-Rodr{\'{\i}}guez}, {Schreiber},
  {van der Plas}, {Williams}, \& {Zurlo}}]{cieza2017}
{Cieza}, L.~A., {Casassus}, S., {P{\'e}rez}, S., {et~al.} 2017, \apjl, 851, L23

\bibitem[{{Cody} {et~al.}(2014){Cody}, {Stauffer}, {Baglin}, {Micela},
  {Rebull}, {Flaccomio}, {Morales-Calder{\'o}n}, {Aigrain}, {Bouvier},
  {Hillenbrand}, {Gutermuth}, {Song}, {Turner}, {Alencar}, {Zwintz},
  {Plavchan}, {Carpenter}, {Findeisen}, {Carey}, {Terebey}, {Hartmann},
  {Calvet}, {Teixeira}, {Vrba}, {Wolk}, {Covey}, {Poppenhaeger}, {G{\"u}nther},
  {Forbrich}, {Whitney}, {Affer}, {Herbst}, {Hora}, {Barrado}, {Holtzman},
  {Marchis}, {Wood}, {Medeiros Guimar{\~a}es}, {Lillo Box}, {Gillen},
  {McQuillan}, {Espaillat}, {Allen}, {D'Alessio}, \& {Favata}}]{cody2014}
{Cody}, A.~M., {Stauffer}, J., {Baglin}, A., {et~al.} 2014, \aj, 147, 82

\bibitem[{{de Boer} {et~al.}(2016){de Boer}, {Salter}, {Benisty}, {Vigan},
  {Boccaletti}, {Pinilla}, {Ginski}, {Juhasz}, {Maire}, {Messina}, {Desidera},
  {Cheetham}, {Girard}, {Wahhaj}, {Langlois}, {Bonnefoy}, {Beuzit}, {Buenzli},
  {Chauvin}, {Dominik}, {Feldt}, {Gratton}, {Hagelberg}, {Isella}, {Janson},
  {Keller}, {Lagrange}, {Lannier}, {Menard}, {Mesa}, {Mouillet}, {Mugrauer},
  {Peretti}, {Perrot}, {Sissa}, {Snik}, {Vogt}, {Zurlo}, \& {SPHERE
  Consortium}}]{deboer2016}
{de Boer}, J., {Salter}, G., {Benisty}, M., {et~al.} 2016, \aap, 595, A114

\bibitem[{{Debes} {et~al.}(2017){Debes}, {Poteet}, {Jang-Condell}, {Gaspar},
  {Hines}, {Kastner}, {Pueyo}, {Rapson}, {Roberge}, {Schneider}, \&
  {Weinberger}}]{debes2017}
{Debes}, J.~H., {Poteet}, C.~A., {Jang-Condell}, H., {et~al.} 2017, \apj, 835,
  205

\bibitem[{{Dipierro} {et~al.}(2015){Dipierro}, {Price}, {Laibe}, {Hirsh},
  {Cerioli}, \& {Lodato}}]{dipierro2015}
{Dipierro}, G., {Price}, D., {Laibe}, G., {et~al.} 2015, \mnras, 453, L73

\bibitem[{{Dipierro} {et~al.}(2018){Dipierro}, {Ricci}, {P{\'e}rez}, {Lodato},
  {Alexander}, {Laibe}, {Andrews}, {Carpenter}, {Chandler}, {Greaves}, {Hall},
  {Henning}, {Kwon}, {Linz}, {Mundy}, {Sargent}, {Tazzari}, {Testi}, \&
  {Wilner}}]{dipierro2018}
{Dipierro}, G., {Ricci}, L., {P{\'e}rez}, L., {et~al.} 2018, \mnras, 475, 5296

\bibitem[{{Dohlen} {et~al.}(2008){Dohlen}, {Langlois}, {Saisse}, {Hill},
  {Origne}, {Jacquet}, {Fabron}, {Blanc}, {Llored}, {Carle}, {Moutou}, {Vigan},
  {Boccaletti}, {Carbillet}, {Mouillet}, \& {Beuzit}}]{dohlen2008}
{Dohlen}, K., {Langlois}, M., {Saisse}, M., {et~al.} 2008, in SPIE Proc., Vol.
  7014

\bibitem[{{Dong} \& {Fung}(2017)}]{dong2017}
{Dong}, R. \& {Fung}, J. 2017, \apj, 835, 146

\bibitem[{{Do{\u{g}}an} {et~al.}(2015){Do{\u{g}}an}, {Nixon}, {King}, \&
  {Price}}]{dougan2015}
{Do{\u{g}}an}, S., {Nixon}, C., {King}, A., \& {Price}, D.~J. 2015, \mnras,
  449, 1251

\bibitem[{{Facchini} {et~al.}(2018){Facchini}, {Juh{\'a}sz}, \&
  {Lodato}}]{facchini2018}
{Facchini}, S., {Juh{\'a}sz}, A., \& {Lodato}, G. 2018, \mnras, 473, 4459

\bibitem[{{Facchini} {et~al.}(2013){Facchini}, {Lodato}, \&
  {Price}}]{facchini2013}
{Facchini}, S., {Lodato}, G., \& {Price}, D.~J. 2013, \mnras, 433, 2142

\bibitem[{{Fedele} {et~al.}(2017){Fedele}, {Carney}, {Hogerheijde}, {Walsh},
  {Miotello}, {Klaassen}, {Bruderer}, {Henning}, \& {van
  Dishoeck}}]{2017A&A...600A..72F}
{Fedele}, D., {Carney}, M., {Hogerheijde}, M.~R., {et~al.} 2017, \aap, 600, A72

\bibitem[{{Fedele} {et~al.}(2018){Fedele}, {Tazzari}, {Booth}, {Testi},
  {Clarke}, {Pascucci}, {Kospal}, {Semenov}, {Bruderer}, {Henning}, \&
  {Teague}}]{fedele2018}
{Fedele}, D., {Tazzari}, M., {Booth}, R., {et~al.} 2018, \aap, 610, A24

\bibitem[{{Flock} {et~al.}(2015){Flock}, {Ruge}, {Dzyurkevich}, {Henning},
  {Klahr}, \& {Wolf}}]{flock2015}
{Flock}, M., {Ruge}, J.~P., {Dzyurkevich}, N., {et~al.} 2015, \aap, 574, A68

\bibitem[{{Foreman-Mackey} {et~al.}(2013){Foreman-Mackey}, {Hogg}, {Lang}, \&
  {Goodman}}]{emcee}
{Foreman-Mackey}, D., {Hogg}, D.~W., {Lang}, D., \& {Goodman}, J. 2013,
  Publications of the Astronomical Society of the Pacific, 125, 306

\bibitem[{Fusco {et~al.}(2006)Fusco, Rousset, Sauvage, Petit, Beuzit, Dohlen,
  Mouillet, Charton, Nicolle, Kasper, Baudoz, \& Puget}]{fusco2006}
Fusco, T., Rousset, G., Sauvage, J.-F., {et~al.} 2006, Opt. Express, 14, 7515

\bibitem[{{Gaia Collaboration} {et~al.}(2018){Gaia Collaboration}, {Brown},
  {Vallenari}, {Prusti}, {de Bruijne}, {Babusiaux}, \&
  {Bailer-Jones}}]{gaia2018}
{Gaia Collaboration}, {Brown}, A.~G.~A., {Vallenari}, A., {et~al.} 2018, ArXiv
  e-prints, arXiv:1804.09365

\bibitem[{{Gallenne} {et~al.}(2015){Gallenne}, {M{\'e}rand}, {Kervella},
  {Monnier}, {Schaefer}, {Baron}, {Breitfelder}, {Le Bouquin}, {Roettenbacher},
  {Gieren}, {Pietrzy{\'n}ski}, {McAlister}, {ten Brummelaar}, {Sturmann},
  {Sturmann}, {Turner}, {Ridgway}, \& {Kraus}}]{gallenne2015}
{Gallenne}, A., {M{\'e}rand}, A., {Kervella}, P., {et~al.} 2015, \aap, 579, A68

\bibitem[{{Garufi} {et~al.}(2013){Garufi}, {Quanz}, {Avenhaus}, {Buenzli},
  {Dominik}, {Meru}, {Meyer}, {Pinilla}, {Schmid}, \& {Wolf}}]{garufi2013}
{Garufi}, A., {Quanz}, S.~P., {Avenhaus}, H., {et~al.} 2013, \aap, 560, A105

\bibitem[{{Hunter}(2007)}]{matplotlib}
{Hunter}, J.~D. 2007, Computing in Science and Engineering, 9, 90

\bibitem[{{Isella} {et~al.}(2016){Isella}, {Guidi}, {Testi}, {Liu}, {Li}, {Li},
  {Weaver}, {Boehler}, {Carperter}, {De Gregorio-Monsalvo}, {Manara}, {Natta},
  {P{\'e}rez}, {Ricci}, {Sargent}, {Tazzari}, \& {Turner}}]{isella2016}
{Isella}, A., {Guidi}, G., {Testi}, L., {et~al.} 2016, Physical Review Letters,
  117, 251101

\bibitem[{{Juh{\'a}sz} {et~al.}(2015){Juh{\'a}sz}, {Benisty}, {Pohl},
  {Dullemond}, {Dominik}, \& {Paardekooper}}]{juhasz2015}
{Juh{\'a}sz}, A., {Benisty}, M., {Pohl}, A., {et~al.} 2015, \mnras, 451, 1147

\bibitem[{{Juh{\'a}sz} \& {Facchini}(2017)}]{juhasz2017}
{Juh{\'a}sz}, A. \& {Facchini}, S. 2017, \mnras, 466, 4053

\bibitem[{{Keppler} {et~al.}(2018){Keppler}, {Benisty}, {M{\"u}ller},
  {Henning}, {van Boekel}, {Cantalloube}, {Ginski}, {van Holstein}, {Maire},
  {Pohl}, {Samland}, {Avenhaus}, {Baudino}, {Boccaletti}, {de Boer},
  {Bonnefoy}, {Chauvin}, {Desidera}, {Langlois}, {Lazzoni}, {Marleau},
  {Mordasini}, {Pawellek}, {Stolker}, {Vigan}, {Zurlo}, {Birnstiel},
  {Brandner}, {Feldt}, {Flock}, {Girard}, {Gratton}, {Hagelberg}, {Isella},
  {Janson}, {Juhasz}, {Kemmer}, {Kral}, {Lagrange}, {Launhardt}, {Matter},
  {M{\'e}nard}, {Milli}, {Molli{\`e}re}, {Olofsson}, {Perez}, {Pinilla},
  {Pinte}, {Quanz}, {Schmidt}, {Udry}, {Wahhaj}, {Williams}, {Buenzli},
  {Cudel}, {Dominik}, {Galicher}, {Kasper}, {Lannier}, {Mesa}, {Mouillet},
  {Peretti}, {Perrot}, {Salter}, {Sissa}, {Wildi}, {Abe}, {Antichi},
  {Augereau}, {Baruffolo}, {Baudoz}, {Bazzon}, {Beuzit}, {Blanchard}, {Brems},
  {Buey}, {De Caprio}, {Carbillet}, {Carle}, {Cascone}, {Cheetham}, {Claudi},
  {Costille}, {Delboulb{\'e}}, {Dohlen}, {Fantinel}, {Feautrier}, {Fusco},
  {Giro}, {Gisler}, {Gluck}, {Gry}, {Hubin}, {Hugot}, {Jaquet}, {Le Mignant},
  {Llored}, {Madec}, {Magnard}, {Martinez}, {Maurel}, {Meyer},
  {Moeller-Nilsson}, {Moulin}, {Mugnier}, {Origne}, {Pavlov}, {Perret},
  {Petit}, {Pragt}, {Puget}, {Rabou}, {Ramos}, {Rigal}, {Rochat}, {Roelfsema},
  {Rousset}, {Roux}, {Salasnich}, {Sauvage}, {Sevin}, {Soenke}, {Stadler},
  {Suarez}, {Turatto}, \& {Weber}}]{keppler2018}
{Keppler}, M., {Benisty}, M., {M{\"u}ller}, A., {et~al.} 2018, ArXiv e-prints,
  arXiv:1806.11568

\bibitem[{{Kraus} {et~al.}(2008){Kraus}, {Ireland}, {Martinache}, \&
  {Lloyd}}]{kraus2008}
{Kraus}, A.~L., {Ireland}, M.~J., {Martinache}, F., \& {Lloyd}, J.~P. 2008,
  \apj, 679, 762

\bibitem[{{Kuhn} {et~al.}(2001){Kuhn}, {Potter}, \& {Parise}}]{kuhn2001}
{Kuhn}, J.~R., {Potter}, D., \& {Parise}, B. 2001, \apjl, 553, L189

\bibitem[{{Langlois} {et~al.}(2014){Langlois}, {Dohlen}, {Vigan}, {Zurlo},
  {Moutou}, {Schmid}, {Mili}, {Beuzit}, {Boccaletti}, {Carle}, {Costille},
  {Dorn}, {Gluck}, {Hubin}, {Feldt}, {Kasper}, {Lizon}, {Madec}, {Le Mignant},
  {Mouillet}, {Puget}, {Sauvage}, \& {Wildi}}]{langlois2014}
{Langlois}, M., {Dohlen}, K., {Vigan}, A., {et~al.} 2014, in SPIE Proc., Vol.
  9147, 1

\bibitem[{{Lavail} {et~al.}(2017){Lavail}, {Kochukhov}, {Hussain}, {Alecian},
  {Herczeg}, \& {Johns-Krull}}]{lavail2017}
{Lavail}, A., {Kochukhov}, O., {Hussain}, G.~A.~J., {et~al.} 2017, \aap, 608,
  A77

\bibitem[{{Lazareff} {et~al.}(2017){Lazareff}, {Berger}, {Kluska}, {Le
  Bouquin}, {Benisty}, {Malbet}, {Koen}, {Pinte}, {Thi}, {Absil}, {Baron},
  {Delboulb{\'e}}, {Duvert}, {Isella}, {Jocou}, {Juhasz}, {Kraus}, {Lachaume},
  {M{\'e}nard}, {Millan-Gabet}, {Monnier}, {Moulin}, {Perraut}, {Rochat},
  {Soulez}, {Tallon}, {Thi{\'e}baut}, {Traub}, \& {Zins}}]{lazareff2017}
{Lazareff}, B., {Berger}, J.-P., {Kluska}, J., {et~al.} 2017, \aap, 599, A85

\bibitem[{{Lodato} \& {Facchini}(2013)}]{2013MNRAS.433.2157L}
{Lodato}, G. \& {Facchini}, S. 2013, \mnras, 433, 2157

\bibitem[{{Loomis} {et~al.}(2017){Loomis}, {{\"O}berg}, {Andrews}, \&
  {MacGregor}}]{loomis2017}
{Loomis}, R.~A., {{\"O}berg}, K.~I., {Andrews}, S.~M., \& {MacGregor}, M.~A.
  2017, \apj, 840, 23

\bibitem[{{Lubow} \& {Martin}(2016)}]{lubow2016}
{Lubow}, S.~H. \& {Martin}, R.~G. 2016, \apj, 817, 30

\bibitem[{{Maire} {et~al.}(2016){Maire}, {Langlois}, {Dohlen}, {Lagrange},
  {Gratton}, {Chauvin}, {Desidera}, {Girard}, {Milli}, {Vigan}, {Zins},
  {Delorme}, {Beuzit}, {Claudi}, {Feldt}, {Mouillet}, {Puget}, {Turatto}, \&
  {Wildi}}]{maire2016}
{Maire}, A.-L., {Langlois}, M., {Dohlen}, K., {et~al.} 2016, in \procspie, Vol.
  9908, Ground-based and Airborne Instrumentation for Astronomy VI, 990834

\bibitem[{{Marino} {et~al.}(2015){Marino}, {Perez}, \& {Casassus}}]{marino2015}
{Marino}, S., {Perez}, S., \& {Casassus}, S. 2015, \apjl, 798, L44

\bibitem[{{Martin} {et~al.}(2016){Martin}, {Lubow}, {Nixon}, \&
  {Armitage}}]{martin2016}
{Martin}, R.~G., {Lubow}, S.~H., {Nixon}, C., \& {Armitage}, P.~J. 2016,
  \mnras, 458, 4345

\bibitem[{{Martinez} {et~al.}(2009){Martinez}, {Dorrer}, {Aller Carpentier},
  {Kasper}, {Boccaletti}, {Dohlen}, \& {Yaitskova}}]{martinez2009}
{Martinez}, P., {Dorrer}, C., {Aller Carpentier}, E., {et~al.} 2009, \aap, 495,
  363

\bibitem[{{Matsakos} \& {K{\"o}nigl}(2017)}]{matsakos2017}
{Matsakos}, T. \& {K{\"o}nigl}, A. 2017, \aj, 153, 60

\bibitem[{{Min} {et~al.}(2017){Min}, {Stolker}, {Dominik}, \&
  {Benisty}}]{min2017}
{Min}, M., {Stolker}, T., {Dominik}, C., \& {Benisty}, M. 2017, \aap, 604, L10

\bibitem[{{Nealon} {et~al.}(2018){Nealon}, {Dipierro}, {Alexander}, {Martin},
  \& {Nixon}}]{nealon2018}
{Nealon}, R., {Dipierro}, G., {Alexander}, R., {Martin}, R.~G., \& {Nixon}, C.
  2018, \mnras, 2159

\bibitem[{{Nixon} {et~al.}(2012){Nixon}, {King}, {Price}, \&
  {Frank}}]{nixon2012b}
{Nixon}, C., {King}, A., {Price}, D., \& {Frank}, J. 2012, \apj, 757, L24

\bibitem[{{Okuzumi} {et~al.}(2016){Okuzumi}, {Momose}, {Sirono}, {Kobayashi},
  \& {Tanaka}}]{okuzumi2016}
{Okuzumi}, S., {Momose}, M., {Sirono}, S.-i., {Kobayashi}, H., \& {Tanaka}, H.
  2016, \apj, 821, 82

\bibitem[{{Owen} \& {Lai}(2017)}]{owen2017}
{Owen}, J.~E. \& {Lai}, D. 2017, \mnras, 469, 2834

\bibitem[{{Pecaut} {et~al.}(2012){Pecaut}, {Mamajek}, \& {Bubar}}]{pecaut2012}
{Pecaut}, M.~J., {Mamajek}, E.~E., \& {Bubar}, E.~J. 2012, \apj, 746, 154

\bibitem[{{Petit} {et~al.}(2014){Petit}, {Sauvage}, {Fusco}, {Sevin}, {Suarez},
  {Costille}, {Vigan}, {Soenke}, {Perret}, {Rochat}, {Barrufolo}, {Salasnich},
  {Beuzit}, {Dohlen}, {Mouillet}, {Puget}, {Wildi}, {Kasper}, {Conan},
  {Kulcs{\'a}r}, \& {Raynaud}}]{petit2014}
{Petit}, C., {Sauvage}, J.-F., {Fusco}, T., {et~al.} 2014, in SPIE Proc., Vol.
  9148, 0

\bibitem[{{Pinilla} {et~al.}(2015){Pinilla}, {de Boer}, {Benisty},
  {Juh{\'a}sz}, {de Juan Ovelar}, {Dominik}, {Avenhaus}, {Birnstiel}, {Girard},
  {Huelamo}, {Isella}, \& {Milli}}]{pinilla2015}
{Pinilla}, P., {de Boer}, J., {Benisty}, M., {et~al.} 2015, \aap, 584, L4

\bibitem[{{Pinilla} {et~al.}(2016){Pinilla}, {Flock}, {Ovelar}, \&
  {Birnstiel}}]{pinilla2016}
{Pinilla}, P., {Flock}, M., {Ovelar}, M.~d.~J., \& {Birnstiel}, T. 2016, \aap,
  596, A81

\bibitem[{{Pinilla} {et~al.}(2018{\natexlab{a}}){Pinilla}, {Natta}, {Manara},
  {Ricci}, {Scholz}, \& {Testi}}]{pinilla2018b}
{Pinilla}, P., {Natta}, A., {Manara}, C.~F., {et~al.} 2018{\natexlab{a}}, \aap,
  615, A95

\bibitem[{{Pinilla} {et~al.}(2017){Pinilla}, {Pohl}, {Stammler}, \&
  {Birnstiel}}]{pinilla2017}
{Pinilla}, P., {Pohl}, A., {Stammler}, S.~M., \& {Birnstiel}, T. 2017, \apj,
  845, 68

\bibitem[{{Pinilla} {et~al.}(2018{\natexlab{b}}){Pinilla}, {Tazzari},
  {Pascucci}, {Youdin}, {Garufi}, {Manara}, {Testi}, {van der Plas},
  {Barenfeld}, {Canovas}, {Cox}, {Hendler}, {P{\'e}rez}, \& {van der
  Marel}}]{pinilla2018a}
{Pinilla}, P., {Tazzari}, M., {Pascucci}, I., {et~al.} 2018{\natexlab{b}},
  \apj, 859, 32

\bibitem[{{Pohl} {et~al.}(2017{\natexlab{a}}){Pohl}, {Benisty}, {Pinilla},
  {Ginski}, {de Boer}, {Avenhaus}, {Henning}, {Zurlo}, {Boccaletti},
  {Augereau}, {Birnstiel}, {Dominik}, {Facchini}, {Fedele}, {Janson},
  {Keppler}, {Kral}, {Langlois}, {Ligi}, {Maire}, {M{\'e}nard}, {Meyer},
  {Pinte}, {Quanz}, {Sauvage}, {Sezestre}, {Stolker}, {Szul{\'a}gyi}, {van
  Boekel}, {van der Plas}, {Villenave}, {Baruffolo}, {Baudoz}, {Le Mignant},
  {Maurel}, {Ramos}, \& {Weber}}]{2017arXiv171006485P}
{Pohl}, A., {Benisty}, M., {Pinilla}, P., {et~al.} 2017{\natexlab{a}}, \apj,
  850, 52

\bibitem[{{Pohl} {et~al.}(2017{\natexlab{b}}){Pohl}, {Sissa}, {Langlois},
  {M{\"u}ller}, {Ginski}, {van Holstein}, {Vigan}, {Mesa}, {Maire}, {Henning},
  {Gratton}, {Olofsson}, {van Boekel}, {Benisty}, {Biller}, {Boccaletti},
  {Chauvin}, {Daemgen}, {de Boer}, {Desidera}, {Dominik}, {Garufi}, {Janson},
  {Kral}, {M{\'e}nard}, {Pinte}, {Stolker}, {Szul{\'a}gyi}, {Zurlo},
  {Bonnefoy}, {Cheetham}, {Cudel}, {Feldt}, {Kasper}, {Lagrange}, {Perrot}, \&
  {Wildi}}]{pohl2017}
{Pohl}, A., {Sissa}, E., {Langlois}, M., {et~al.} 2017{\natexlab{b}}, \aap,
  605, A34

\bibitem[{{Preibisch} {et~al.}(2002){Preibisch}, {Brown}, {Bridges},
  {Guenther}, \& {Zinnecker}}]{preibisch2002}
{Preibisch}, T., {Brown}, A.~G.~A., {Bridges}, T., {Guenther}, E., \&
  {Zinnecker}, H. 2002, \aj, 124, 404

\bibitem[{{Price} {et~al.}(2018){Price}, {Cuello}, {Pinte}, {Mentiplay},
  {Casassus}, {Christiaens}, {Kennedy}, {Cuadra}, {Sebastian Perez}, {Marino},
  {Armitage}, {Zurlo}, {Juhasz}, {Ragusa}, {Laibe}, \& {Lodato}}]{price2018}
{Price}, D.~J., {Cuello}, N., {Pinte}, C., {et~al.} 2018, \mnras, 477, 1270

\bibitem[{{Price} {et~al.}(2017){Price}, {Wurster}, {Nixon}, {Tricco},
  {Toupin}, {Pettitt}, {Chan}, {Laibe}, {Glover}, {Dobbs}, {Nealon}, {Liptai},
  {Worpel}, {Bonnerot}, {Dipierro}, {Ragusa}, {Federrath}, {Iaconi},
  {Reichardt}, {Forgan}, {Hutchison}, {Constantino}, {Ayliffe}, {Mentiplay},
  {Hirsh}, \& {Lodato}}]{2017arXiv170203930P}
{Price}, D.~J., {Wurster}, J., {Nixon}, C., {et~al.} 2017, ArXiv e-prints
  [\eprint[arXiv]{1702.03930}]

\bibitem[{{Rigliaco} {et~al.}(2015){Rigliaco}, {Pascucci}, {Duchene},
  {Edwards}, {Ardila}, {Grady}, {Mendigut{\'{\i}}a}, {Montesinos}, {Mulders},
  {Najita}, {Carpenter}, {Furlan}, {Gorti}, {Meijerink}, \&
  {Meyer}}]{rigliaco2015}
{Rigliaco}, E., {Pascucci}, I., {Duchene}, G., {et~al.} 2015, \apj, 801, 31

\bibitem[{{Rosenfeld} {et~al.}(2012){Rosenfeld}, {Qi}, {Andrews}, {Wilner},
  {Corder}, {Dullemond}, {Lin}, {Hughes}, {D'Alessio}, \& {Ho}}]{rosenfeld2012}
{Rosenfeld}, K.~A., {Qi}, C., {Andrews}, S.~M., {et~al.} 2012, \apj, 757, 129

\bibitem[{{Rosotti} {et~al.}(2016){Rosotti}, {Juhasz}, {Booth}, \&
  {Clarke}}]{rosotti2016}
{Rosotti}, G.~P., {Juhasz}, A., {Booth}, R.~A., \& {Clarke}, C.~J. 2016,
  \mnras, 459, 2790

\bibitem[{{Salyk} {et~al.}(2013){Salyk}, {Herczeg}, {Brown}, {Blake},
  {Pontoppidan}, \& {van Dishoeck}}]{salyk2013}
{Salyk}, C., {Herczeg}, G.~J., {Brown}, J.~M., {et~al.} 2013, \apj, 769, 21

\bibitem[{{Sauvage} {et~al.}(2014){Sauvage}, {Fusco}, {Petit}, {Mouillet},
  {Dohlen}, {Costille}, {Beuzit}, {Baruffolo}, {Kasper}, {Suarez Valles},
  {Downing}, {Feautrier}, {Mugnier}, \& {Baudoz}}]{sauvage2014}
{Sauvage}, J., {Fusco}, T., {Petit}, C., {et~al.} 2014, in SPIE Proc., Vol.
  9148

\bibitem[{{Schmid} {et~al.}(2006){Schmid}, {Joos}, \& {Tschan}}]{schmid2006}
{Schmid}, H.~M., {Joos}, F., \& {Tschan}, D. 2006, \aap, 452, 657

\bibitem[{{Schneider} {et~al.}(2018){Schneider}, {Manara}, {Facchini},
  {G{\"u}nther}, {Herczeg}, {Fedele}, \& {Teixeira}}]{schneider2018}
{Schneider}, P.~C., {Manara}, C.~F., {Facchini}, S., {et~al.} 2018, \aap, 614,
  A108

\bibitem[{{Stolker} {et~al.}(2016{\natexlab{a}}){Stolker}, {Dominik},
  {Avenhaus}, {Min}, {de Boer}, {Ginski}, {Schmid}, {Juhasz}, {Bazzon},
  {Waters}, {Garufi}, {Augereau}, {Benisty}, {Boccaletti}, {Henning},
  {Langlois}, {Maire}, {M{\'e}nard}, {Meyer}, {Pinte}, {Quanz}, {Thalmann},
  {Beuzit}, {Carbillet}, {Costille}, {Dohlen}, {Feldt}, {Gisler}, {Mouillet},
  {Pavlov}, {Perret}, {Petit}, {Pragt}, {Rochat}, {Roelfsema}, {Salasnich},
  {Soenke}, \& {Wildi}}]{stolker2016}
{Stolker}, T., {Dominik}, C., {Avenhaus}, H., {et~al.} 2016{\natexlab{a}},
  \aap, 595, A113

\bibitem[{{Stolker} {et~al.}(2016{\natexlab{b}}){Stolker}, {Dominik}, {Min},
  {Garufi}, {Mulders}, \& {Avenhaus}}]{stolker2016b}
{Stolker}, T., {Dominik}, C., {Min}, M., {et~al.} 2016{\natexlab{b}}, \aap,
  596, A70

\bibitem[{{Terquem} \& {Papaloizou}(2002)}]{terquem2002}
{Terquem}, C. \& {Papaloizou}, J. C.~B. 2002, \mnras, 332, L39

\bibitem[{{Teyssandier} {et~al.}(2013){Teyssandier}, {Terquem}, \&
  {Papaloizou}}]{teyssandier2013}
{Teyssandier}, J., {Terquem}, C., \& {Papaloizou}, J. C.~B. 2013, \mnras, 428,
  658

\bibitem[{{van Boekel} {et~al.}(2017){van Boekel}, {Henning}, {Menu}, {de
  Boer}, {Langlois}, {M{\"u}ller}, {Avenhaus}, {Boccaletti}, {Schmid},
  {Thalmann}, {Benisty}, {Dominik}, {Ginski}, {Girard}, {Gisler}, {Lobo Gomes},
  {Menard}, {Min}, {Pavlov}, {Pohl}, {Quanz}, {Rabou}, {Roelfsema}, {Sauvage},
  {Teague}, {Wildi}, \& {Zurlo}}]{vanboekel2017}
{van Boekel}, R., {Henning}, T., {Menu}, J., {et~al.} 2017, \apj, 837, 132

\bibitem[{{van der Marel} {et~al.}(2013){van der Marel}, {van Dishoeck},
  {Bruderer}, {Birnstiel}, {Pinilla}, {Dullemond}, {van Kempen}, {Schmalzl},
  {Brown}, {Herczeg}, {Mathews}, \& {Geers}}]{2013Sci...340.1199V}
{van der Marel}, N., {van Dishoeck}, E.~F., {Bruderer}, S., {et~al.} 2013,
  Science, 340, 1199

\bibitem[{{Walsh} {et~al.}(2017){Walsh}, {Daley}, {Facchini}, \&
  {Juh{\'a}sz}}]{walsh2017}
{Walsh}, C., {Daley}, C., {Facchini}, S., \& {Juh{\'a}sz}, A. 2017, \aap, 607,
  A114

\bibitem[{{Weingartner} \& {Draine}(2001)}]{weingartner_2001}
{Weingartner}, J.~C. \& {Draine}, B.~T. 2001, \apj, 548, 296

\bibitem[{{Xiang-Gruess} \& {Papaloizou}(2013)}]{xg2013}
{Xiang-Gruess}, M. \& {Papaloizou}, J.~C.~B. 2013, \mnras, 431, 1320

\bibitem[{{Zhang} {et~al.}(2015){Zhang}, {Blake}, \& {Bergin}}]{zhang2015}
{Zhang}, K., {Blake}, G.~A., \& {Bergin}, E.~A. 2015, \apjl, 806, L7

\end{thebibliography}

\appendix
\section{$U_{\phi}$ $J$-band images}
\begin{figure*}
\center
\begin{tabular}{ccc}
\includegraphics[width=0.33\textwidth]{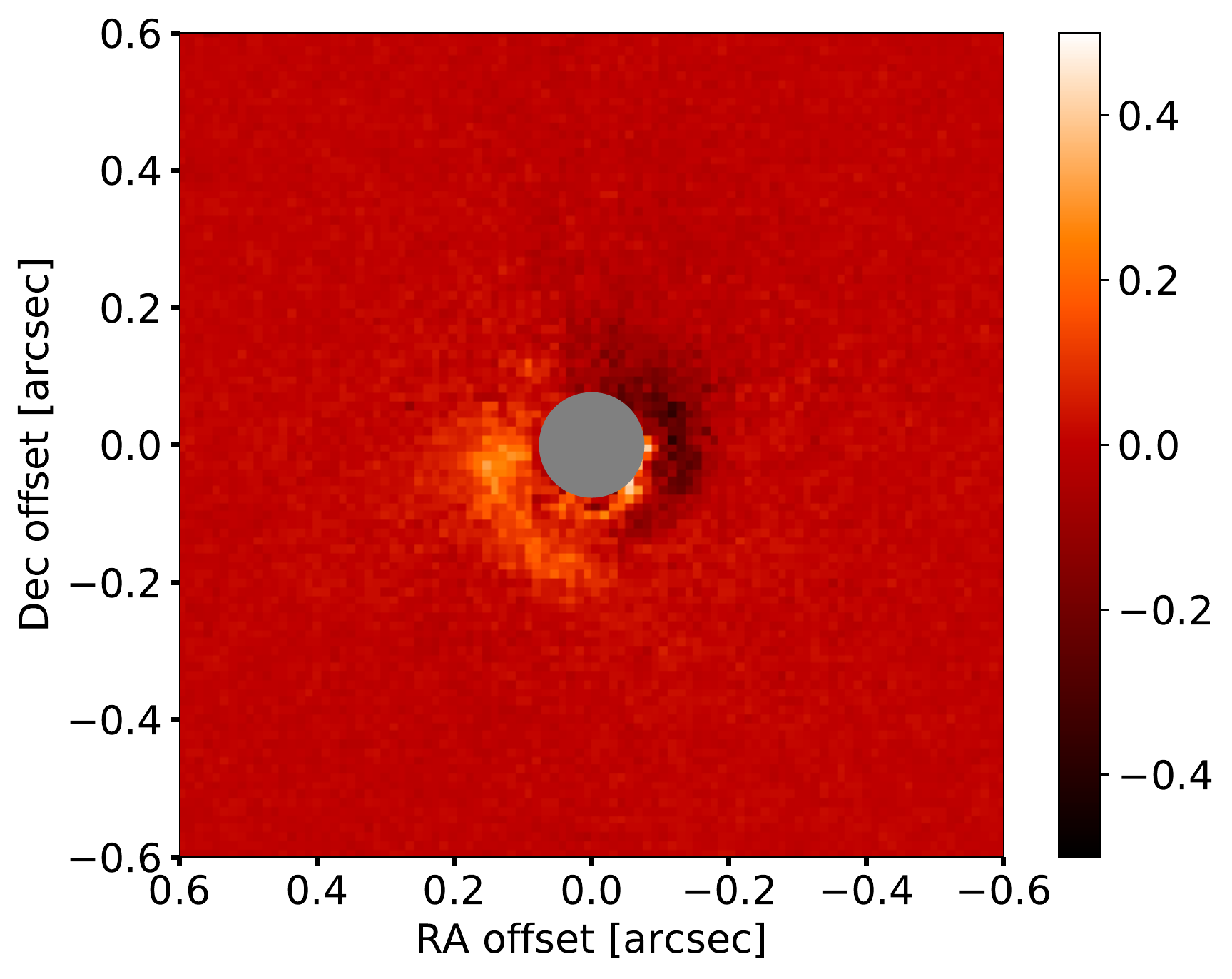} & 
\includegraphics[width=0.33\textwidth]{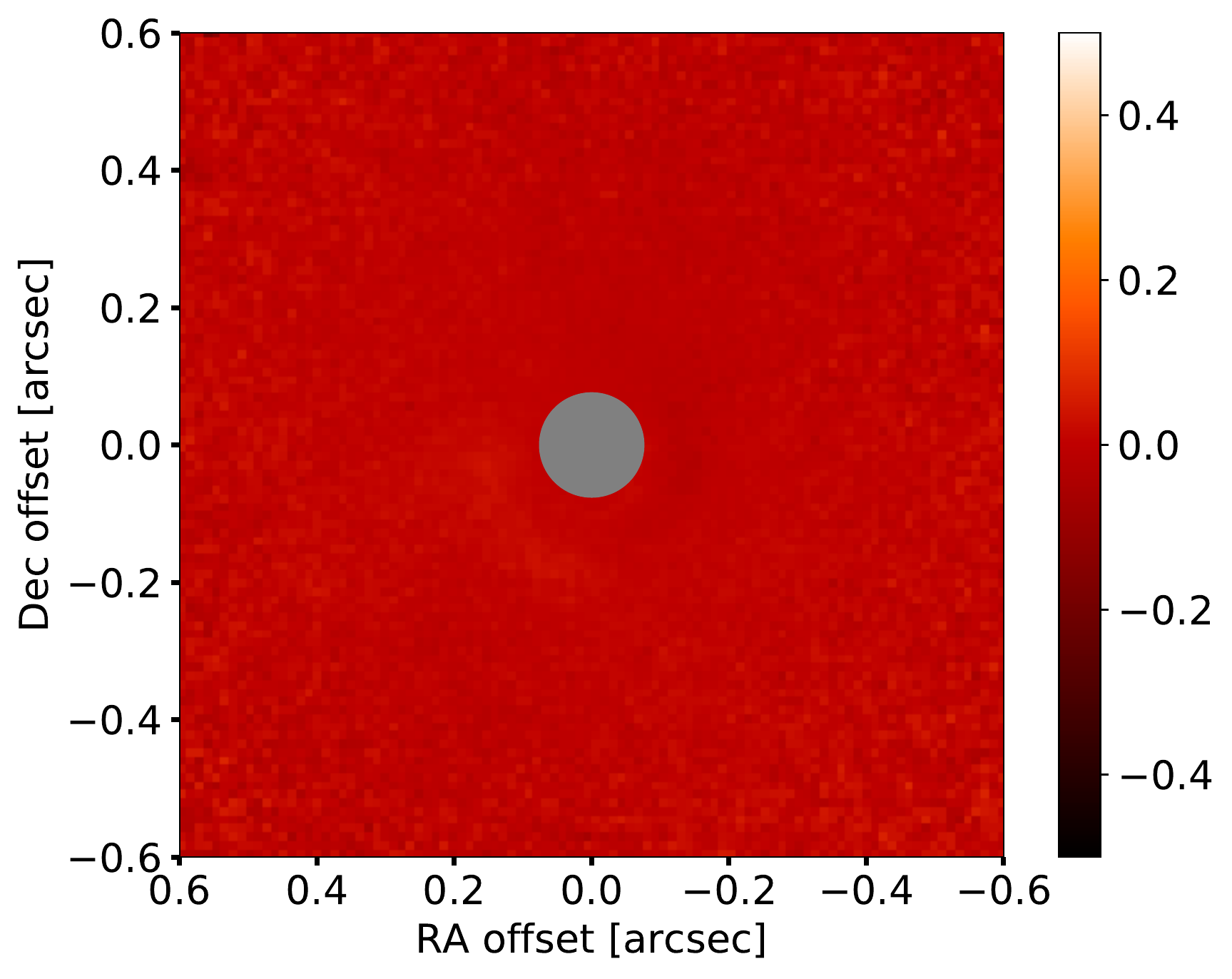} & 
\includegraphics[width=0.33\textwidth]{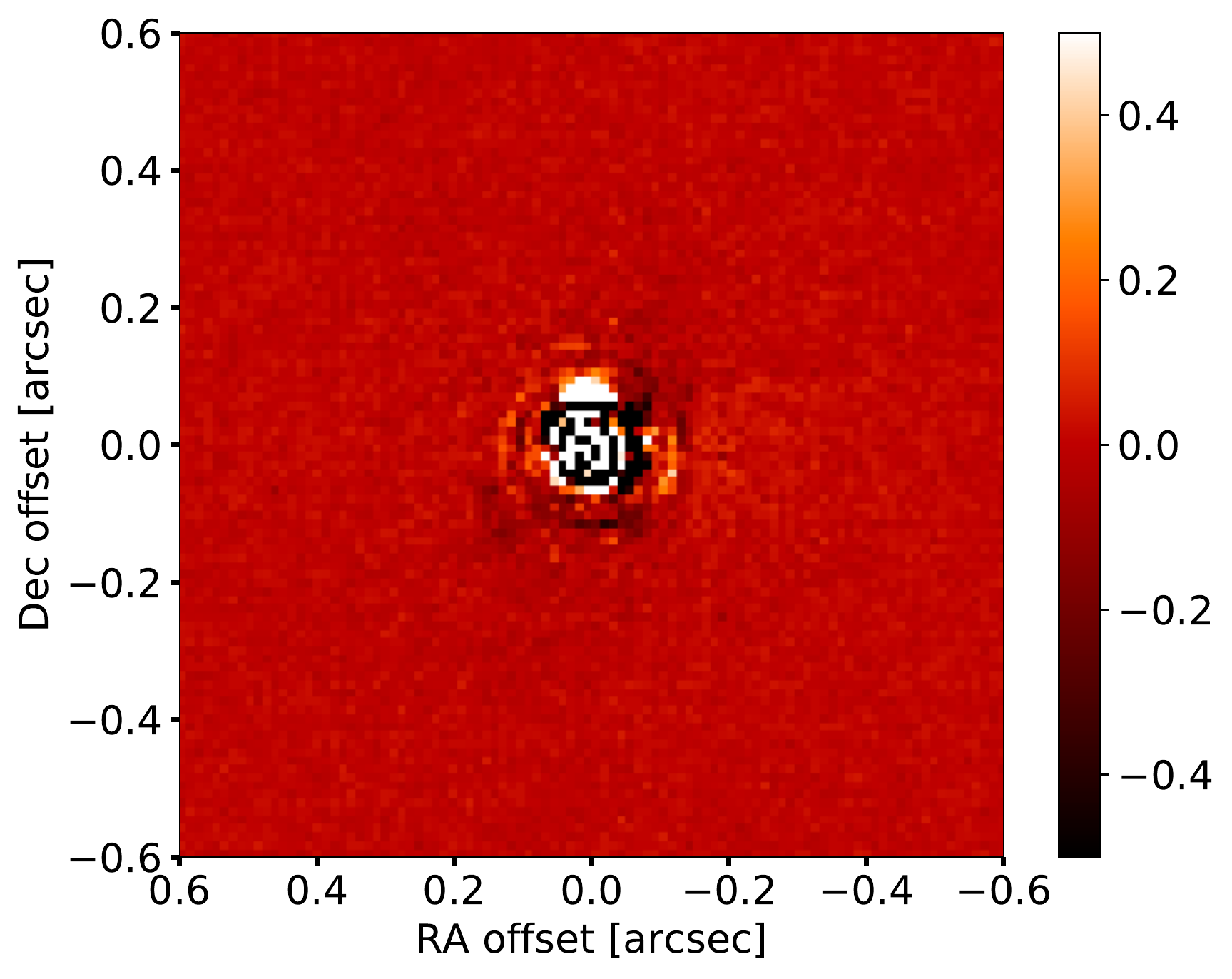}
\end{tabular}
\caption{$J$-band $U_{\phi}$ images. Left: Coronagraphic $U_{\phi}$ image; Middle: Same but r$^{2}$-scaled; Right: Non-coronagraphic $U_{\phi}$ image. The $U_{\phi}$ images are shown with the same dynamical range as the corresponding $Q_{\phi}$ images in Fig.\,\ref{fig:obsQphi}.}
\label{fig:obsUphi}
\end{figure*}

Figure \ref{fig:obsUphi} presents the $U_{\phi}$ images, for the two sets of scattered light observations presented in this paper. 

\section{Moment 1 CO map}
\begin{figure*}
\center
\begin{tabular}{ccc}
\includegraphics[width=0.33\textwidth]{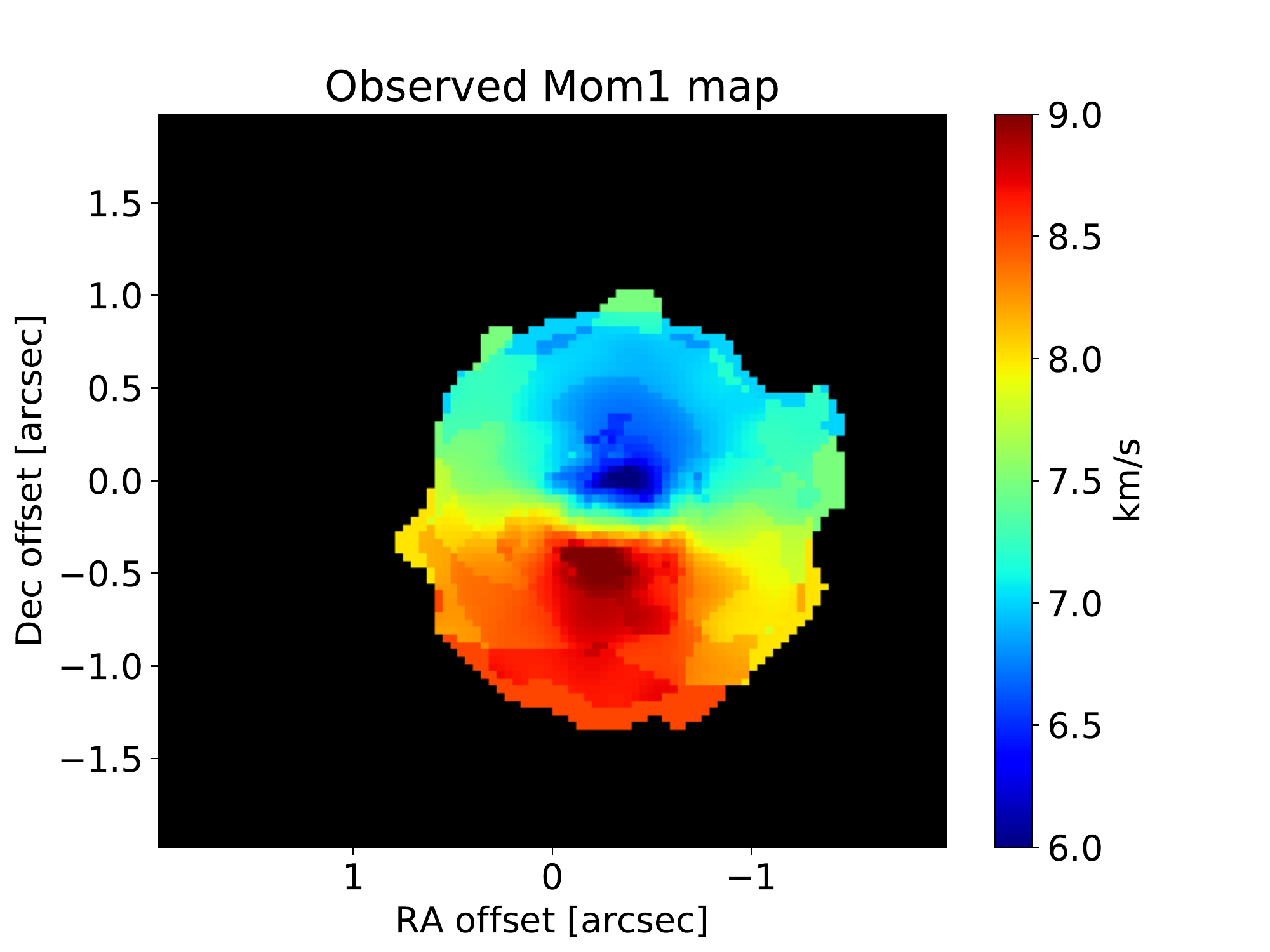} & 
\includegraphics[width=0.33\textwidth]{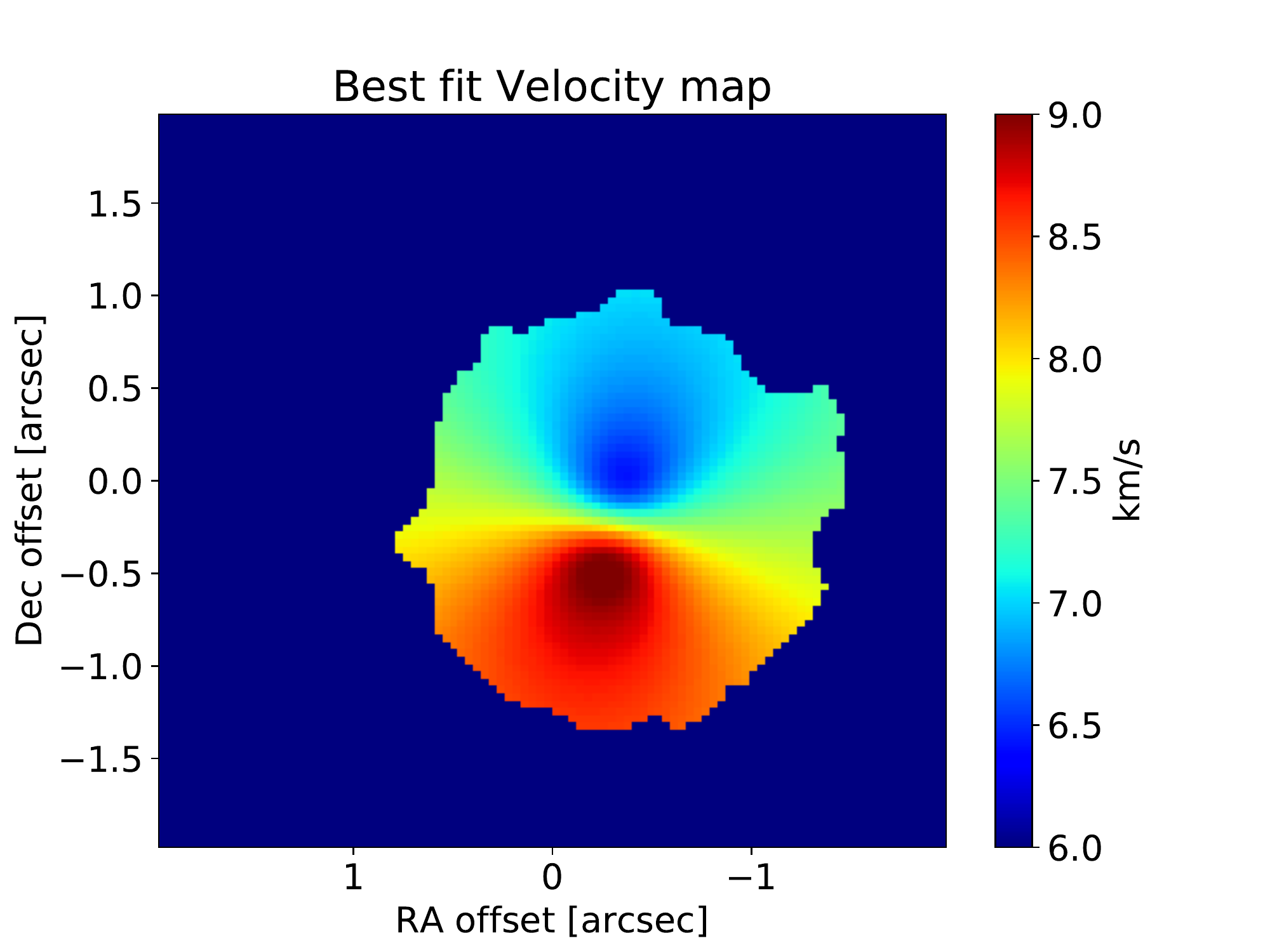} & 
\includegraphics[width=0.33\textwidth]{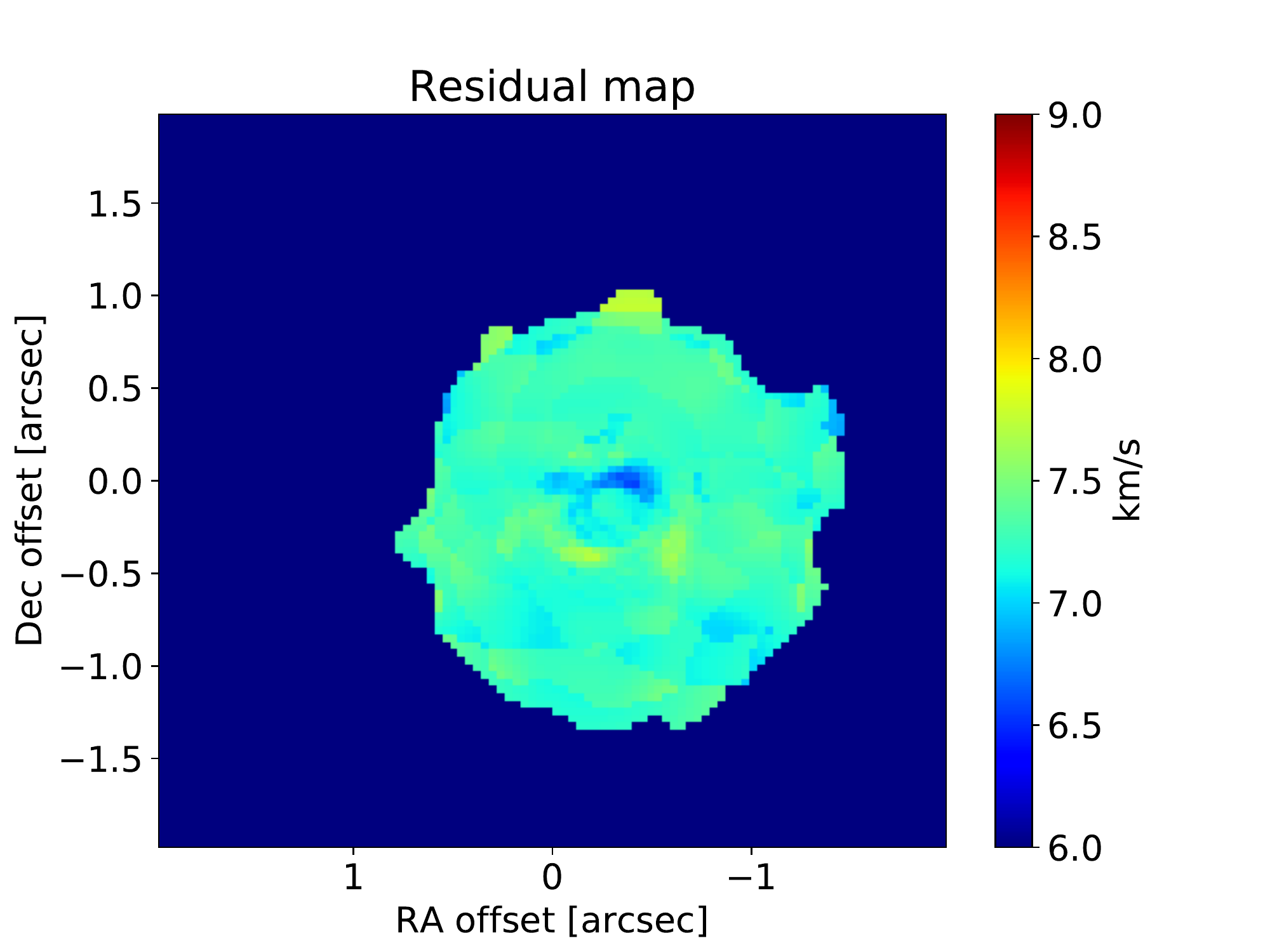}
\end{tabular}
\caption{
Left: Moment 1 map derived from the ALMA data of \citet{barenfeld2016}. Middle: Velocity map from our best-fit model. Right: Residuals.} 
\label{fig:mom1}
\end{figure*}

\begin{figure*}
\center
\includegraphics[width=0.8\textwidth]{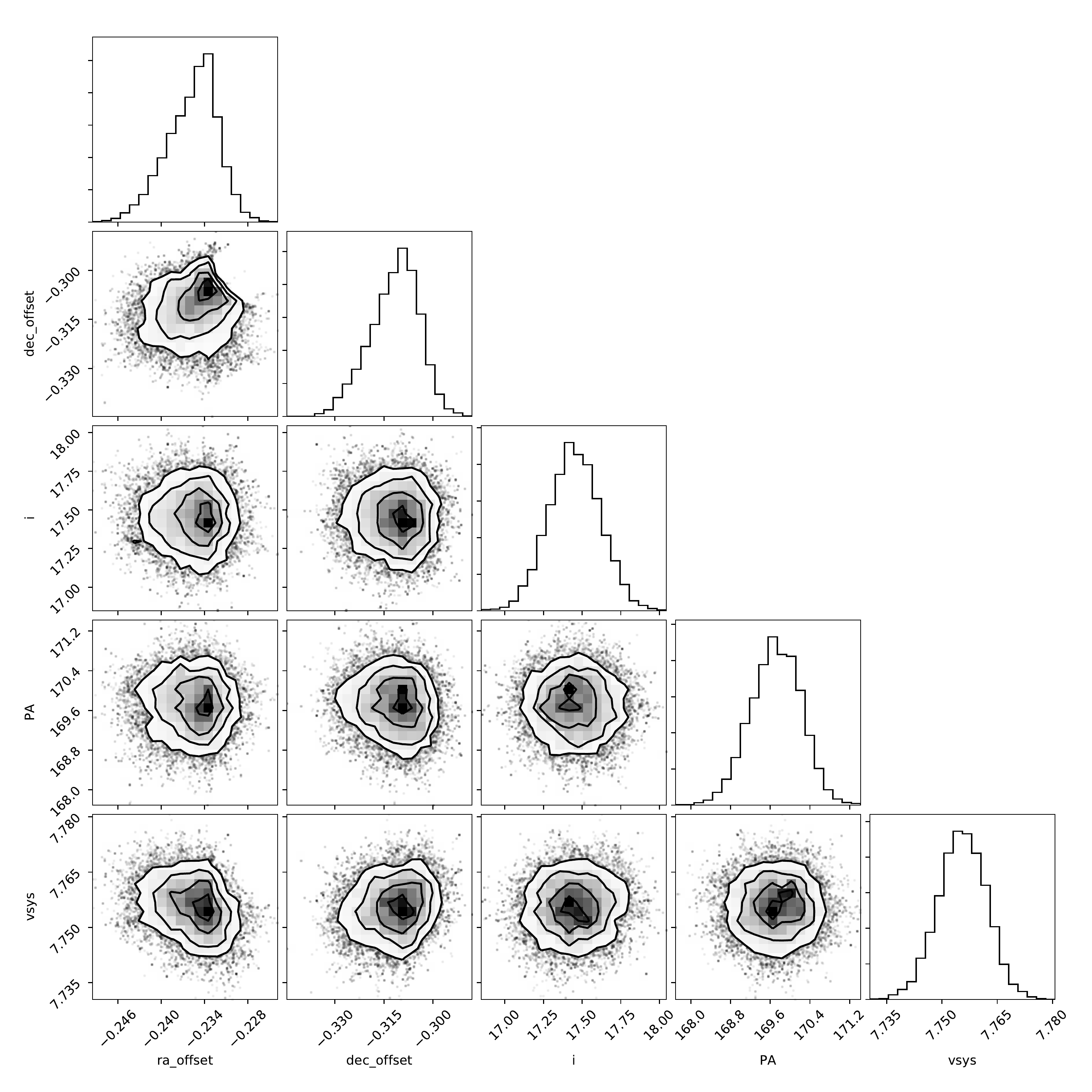} 
\caption{MCMC chains for the fit of the CO Moment 1 map using a Keplerian disk model.}
\label{fig:momCOFit}
\end{figure*}

Figures \ref{fig:mom1} and \ref{fig:momCOFit} show the Moment 1 map, the best-fit velocity map using a razor-thin Keplerian disk model, the residuals, and the MCMC chains. 

\section{SPH snapshot}

\begin{figure}
\centering
\includegraphics[width=0.49\textwidth]{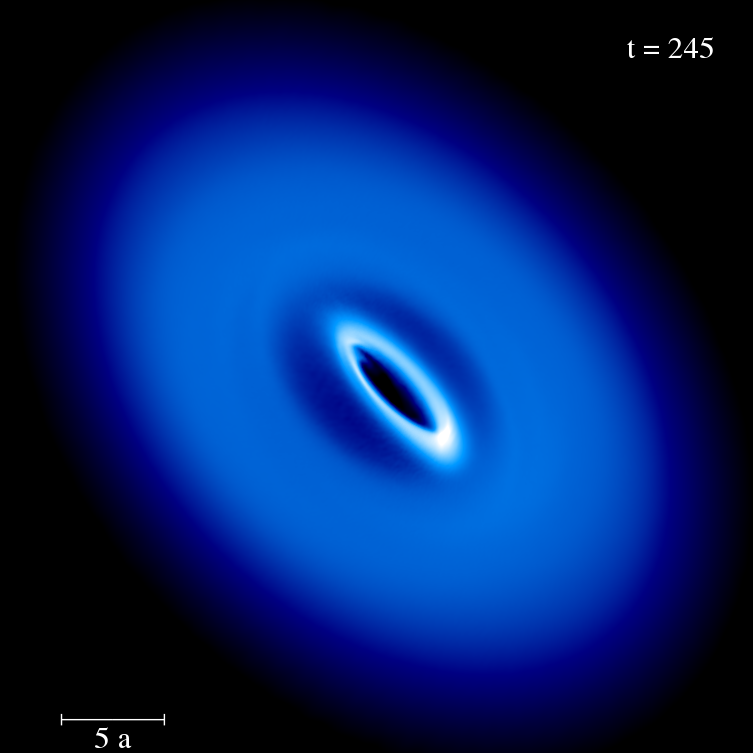}
\caption{Snapshot of the hydrodynamical simulation considered in this paper, corresponding to 245 binary orbits.} 
\label{fig:sph}
\end{figure}

Figure\,\ref{fig:sph} shows the surface density of the SPH simulation considered in the paper. The details of the simulation are presented in \citet{facchini2018}. The physical model consists of a central equal mass binary misaligned to the outer disk by $60^\circ$. The gravitational torque breaks the disk into two distinct annuli. In this particular snapshot, taken after 245 binary orbits, the mutual misalignment between inner and outer disk is $\sim30^\circ$.

\section{ZIMPOL optical image}
\begin{figure*}
\center
\begin{tabular}{cc}
\includegraphics[width=0.4\textwidth]{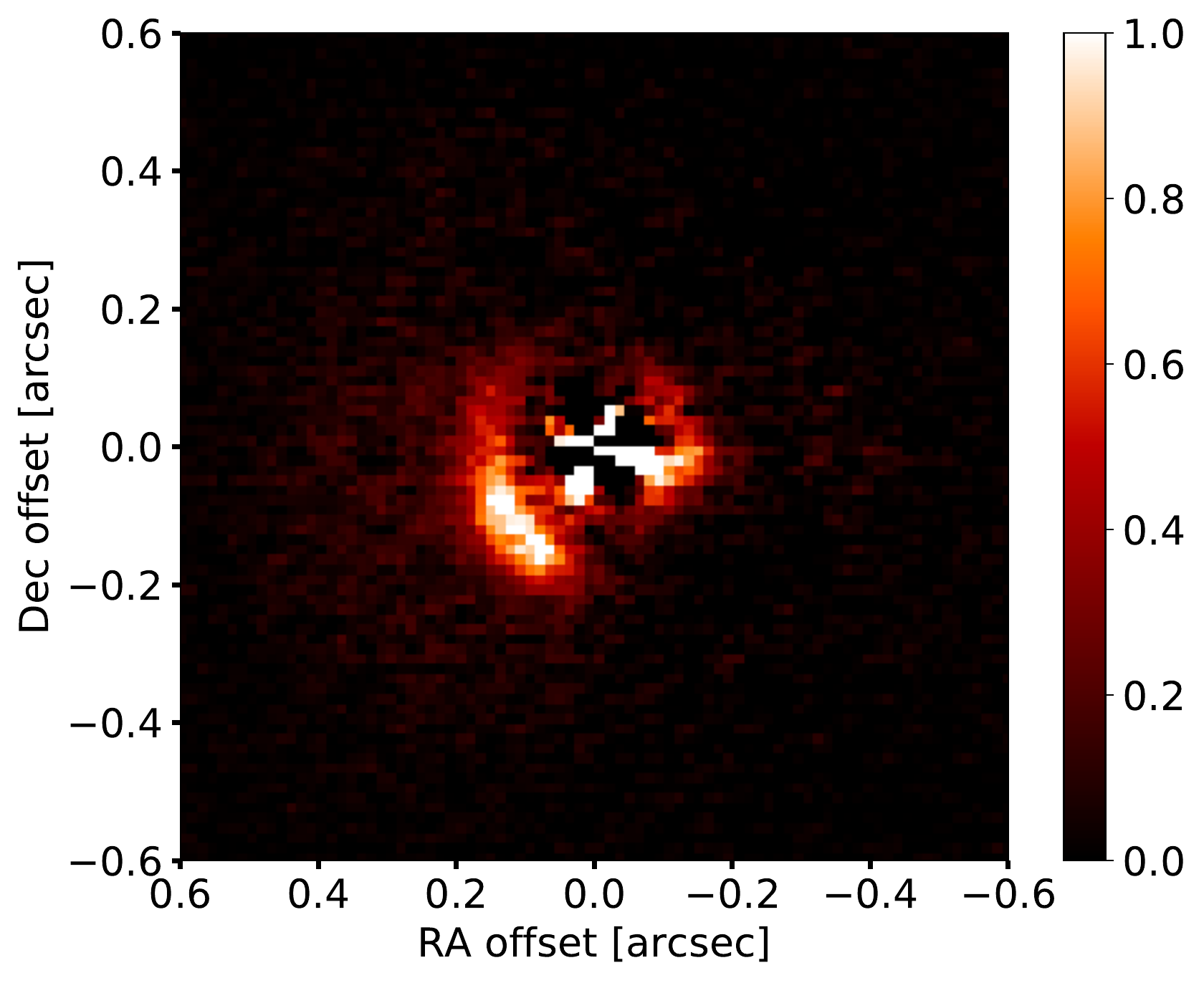} &
\includegraphics[width=0.4\textwidth]{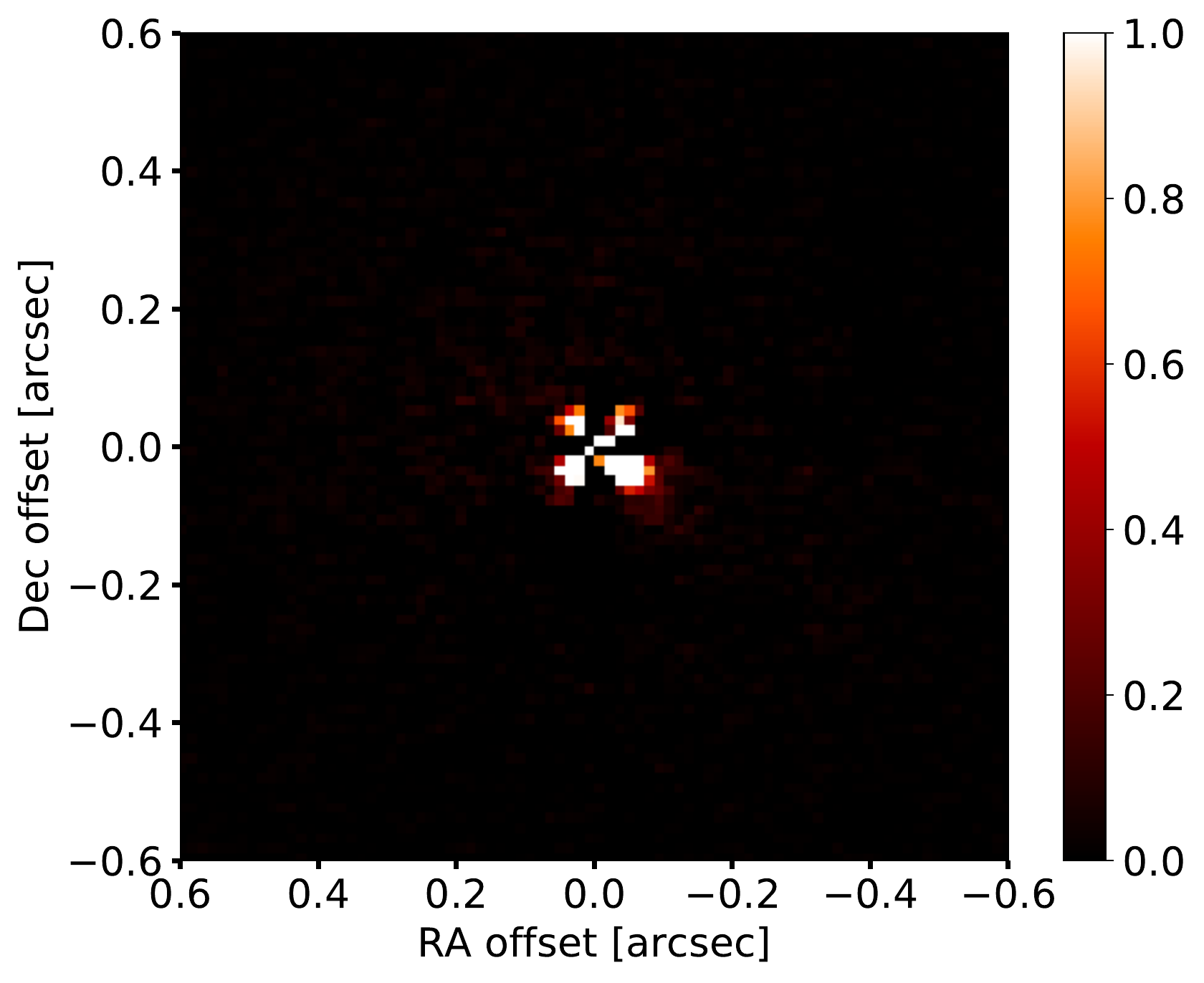}
\end{tabular}
\caption{Optical ZIMPOL $Q_{\phi}$ and $U_{\phi}$ images (left, and right, respectively) obtained on 2015 June 10. }
\label{fig:zim}
\end{figure*}

Figure \ref{fig:zim} shows an archival ZIMPOL observation from 2015 June 10.  The data were reduced following \citet{deboer2016} and as described in Sect.\,\ref{sec:datared}, and support that the shadows do not move very quickly, as their locations are similar between 2015 and 2016. 

\label{lastpage}
\end{document}